\documentclass[aps,prb,twocolumn,floatfix,superscriptaddress,10pt]{revtex4-1}

\usepackage{amsmath}
\usepackage{amsfonts}
\usepackage{graphicx}
\usepackage{verbatim} 
\usepackage{color}
\usepackage{subfigure}
\usepackage{hyperref}
\usepackage{braket}

\begin{document}

\title{The Thouless theorem for matrix product states and subsequent post-density matrix renormalization group methods}
\author{Sebastian Wouters}
\email{sebastianwouters [at] gmail.com}
\affiliation{Center for Molecular Modeling, Ghent University, Technologiepark 903, 9052 Zwijnaarde, Belgium}
\author{Naoki Nakatani}
\affiliation{Department of Chemistry, Princeton University, Frick Chemistry Laboratory, Princeton, NJ 08544, USA}
\author{Dimitri Van Neck}
\affiliation{Center for Molecular Modeling, Ghent University, Technologiepark 903, 9052 Zwijnaarde, Belgium}
\author{Garnet Kin-Lic Chan}
\affiliation{Department of Chemistry, Princeton University, Frick Chemistry Laboratory, Princeton, NJ 08544, USA}

\begin{abstract}
The similarities between Hartree-Fock (HF) theory and the density-matrix renormalization group (DMRG) are explored. Both methods can be formulated as the variational optimization of a wave-function ansatz. Linearization of the time-dependent variational principle near a variational minimum allows to derive the random phase approximation (RPA). We show that the non-redundant parametrization of the matrix product state (MPS) tangent space [J. Haegeman \textit{et al.}, Phys. Rev. Lett. \textbf{107}, 070601 (2011)] leads to the Thouless theorem for MPS, i.e. an explicit non-redundant parametrization of the entire MPS manifold, starting from a specific MPS reference. Excitation operators are identified, which extends the analogy between HF and DMRG to the Tamm-Dancoff approximation (TDA), the configuration interaction (CI) expansion, and coupled cluster theory. For a small one-dimensional Hubbard chain, we use a CI-MPS ansatz with single and double excitations to improve on the ground state and to calculate low-lying excitation energies. For a symmetry-broken ground state of this model, we show that RPA-MPS allows to retrieve the Goldstone mode. We also discuss calculations of the RPA-MPS correlation energy. With the long-range quantum chemical Pariser-Parr-Pople Hamiltonian, low-lying TDA-MPS and RPA-MPS excitation energies for polyenes are obtained.
\end{abstract}

\maketitle
\section{Introduction}
The standard classification of quantum ground states dates back to Landau.\cite{Landau1,*Landau2} Mean-field theory is used to describe a state, and a phase transition is marked by the breaking of a symmetry. Particle-conserving mean-field theory for fermions is Hartree-Fock (HF) theory.\cite{Hartree1,*Hartree2,*SlaterSCF,*FockHF} In HF theory, the exact ground state is approximated by a Slater determinant (SD),\cite{SlaterDeterminant} and the energy of a Hamiltonian is minimized within this variational ansatz space. To obtain excited states or a more accurate description of the ground state, post-HF (post mean-field) methods\cite{helgaker2, *helgaker:52} can be carried out such as the Tamm-Dancoff approximation (TDA),\cite{Tamm, *PhysRev.78.382} the random-phase approximation (RPA),\cite{PhysRev.92.609} M\o{}ller-Plesset perturbation theory,\cite{PhysRev.46.618} the configuration interaction (CI) expansion,\cite{SlaterDeterminant,PhysRev.36.1121} and coupled cluster (CC) theory.\cite{Coester1958421,*Coester1960477,*cizek:4256}\\
Within the framework of second quantization,\cite{DiracSecondQuantization,FockSpace,DimitriBook} the reference SD obtains a simple product form when the canonical HF orbitals are used to construct the Fock space. Occupied-virtual (OV) excitation operators allow to connect the reference SD to post-HF wavefunction ansatzes. The Thouless theorem gives a non-redundant parametrization to generate \textit{all} possible SDs from any given SD reference, by means of its OV excitation operators.\cite{PhysRevA.22.2362,*PhysRevA.24.673,Thouless1960225,*Thouless196178}\\
Recently, a new way to understand the qualitative structure of quantum many-body states has appeared, whereby the state is approximated by a tensor network, i.e. a contracted product of tensors where each tensor represents a local degree of freedom. These ansatzes are efficient representations of low-energy states because they capture the boundary law for the entanglement entropy. In one dimension, the tensor network is known as a matrix product state (MPS). The MPS is the wavefunction ansatz for the density-matrix renormalization group (DMRG) algorithm.\cite{PhysRevLett.69.2863, *PhysRevB.48.10345, *PhysRevLett.75.3537, *PhysRevB.55.2164, *PhysRevB.78.035124, *1742-5468-2007-08-P08024}\\
DMRG can capture states beyond the realm of Landau (or mean-field) theory, i.e. states with topological order.\cite{TopoFirst, *PhysRevB.83.035107, *PhysRevB.83.075103, *PhysRevB.83.075102, *PhysRevB.84.165139, *PhysRevB.86.245305} DMRG has also been shown to be a powerful method to treat the static correlation problem in electronic structure theory.\cite{white:4127, *chan:4462, wouters, *sharma:124121, chan:annurevphys} Static correlation arises when a state consists of several significant SD contributions, which HF theory is of course unable to deal with, because a single SD does not describe the qualitative structure of the targeted state. Post-HF methods, which start from a single SD reference, have difficulty building in large static correlation a posteriori. In these situations DMRG has provided a new ability to access the electronic structure. The analogue of static correlation for DMRG is a quantum critical system, which introduces corrections to the entanglement boundary law which cannot be captured by DMRG.\\
DMRG can be interpreted as a mean-field theory in the sites, which is analogous to HF, which is a mean-field theory in the particles.\cite{B805292C, 2011arXiv1103.2155K, *2013APS..MARJ24006K} Therefore, it is natural to search for extensions to DMRG that are analogous to post-HF methods: post-DMRG methods. One example is linear response theory. Time-dependent HF theory is obtained by using an SD ansatz in the time-dependent variational principle (TDVP).\cite{PhysRevA.22.2362, *PhysRevA.24.673, PSP:2040328, *RevModPhys.44.602, *Kerman1976332, *Kramer} Time-dependent DMRG (which stays within the MPS ansatz space) is similarly obtained by using an MPS ansatz in the TDVP.\cite{PhysRevLett.107.070601,2006cond.mat.12480U,2011arXiv1103.2155K, *2013APS..MARJ24006K, juthothesis} RPA, or linear response theory for HF, is obtained by linearizing the time-dependent HF equations in the vicinity of a variational mimimum.\cite{Zyrianov,*PhysRev.107.450,*PhysRev.107.1631} Equivalently, the RPA equations can be derived from an equation of motion (EOM) approach with excitation operators.\cite{PhysRev.106.372,*PhysRev.108.507,*PhysRev.112.1900, *PhysRev.115.786,*GoldStoneRPAEq} RPA yields a mean-field description of quasi-particle excitations. Linear response theory for DMRG was first derived by Dorando \textit{et al.},\cite{dorando:184111} and was later recast as RPA for MPS.\cite{2011arXiv1103.2155K, *2013APS..MARJ24006K, juthothesis}\\
In this work, we construct a more complete analogue of the mean-field framework, which allows us to define a full set of post-DMRG methods. We give a non-redundant parametrization of the entire MPS manifold, starting from a specific MPS reference. This is the analogue of the Thouless theorem for HF. We identify the excitation operators of the Thouless theorem. These excitation operators allow for a complete rederivation of RPA for MPS by means of the EOM, in complete analogy with HF. All other results, such as an improvement of the ground state theory by the fluctuation-dissipation theorem, follow. With these excitation operators, we can define the analogues of other post-HF methods for MPS, such as CC and CI.\\
For a small one-dimensional Hubbard chain, we use a numerical CI-MPS ansatz with single and double excitations to improve on the ground state and to calculate low-lying excitation energies. For a symmetry-broken ground state of this model, we show that RPA-MPS allows to retrieve the Goldstone mode. We also discuss calculations of the RPA-MPS correlation energy. With the long-range quantum chemical Pariser-Parr-Pople (PPP) Hamiltonian, low-lying TDA-MPS and RPA-MPS excitation energies for polyenes are also obtained.

\section{HF mean-field theory}
This section provides a brief introduction to the variational principles, HF mean-field theory, the Thouless theorem, and post-HF methods. It focusses on the topics for which a DMRG analogue will be constructed in this paper. For readers familiar with HF, this section can be a good guideline to understand our post-DMRG discussion.

\subsection{Variational principles}
Because the Hilbert space increases exponentially with system size, a variational wave-function ansatz $\ket{\Phi(\mathbf{z})}$ with parametrization $\mathbf{z}$ is often used to make calculations feasible. In order to minimize the energy functional
\begin{equation}
E(\mathbf{z},\overline{\mathbf{z}}) = \frac{\braket{ \overline{\Phi} \mid \hat{H} \mid \Phi}}{\braket{\overline{\Phi} \mid \Phi}}
\end{equation}
to approximate ground states, the time-independent variational principle (TIVP) $\frac{\delta \mathcal{L}}{\delta \overline{z}} = 0$ can be employed, where the Lagrangian is\cite{B805292C}
\begin{equation}
\mathcal{L} = \braket{\overline{\Phi} \mid \hat{H} \mid \Phi} - \lambda \left( \braket{\overline{\Phi} \mid \Phi} - 1 \right). \label{TIVP}
\end{equation}
The overline denotes complex conjugation. This yields the time-independent self-consistent field (SCF) equations. To approximate time evolution, the time-dependent variational principle (TDVP) $\frac{\delta \mathcal{S}}{\delta \overline{z}} = 0$ can be employed, where the action is\cite{PhysRevA.22.2362,*PhysRevA.24.673, PSP:2040328, *RevModPhys.44.602, *Kerman1976332, *Kramer, 2006cond.mat.12480U}
\begin{equation}
\mathcal{S} = \int\limits_{t_1}^{t_2} dt \left( \frac{i \hbar}{2} \braket{\overline{\Phi} \mid \dot{\Phi}} - \frac{i \hbar}{2} \braket{\dot{\overline{\Phi}} \mid \Phi} -  \braket{\overline{\Phi} \mid \hat{H} \mid \Phi}  \right).
\end{equation}
The dot denotes time derivation. This yields the time-dependent SCF equations.

\subsection{The Slater determinant}
From a given single-particle basis, any other single-particle basis can be constructed by a unitary transformation: $\hat{a}^{\dagger}_j = \hat{b}_k^{\dagger} U^k_{~j}$. Second quantization is used to denote the single-particle states,\cite{DiracSecondQuantization,FockSpace,DimitriBook} and the summation convention is used for double indices. An $N$-particle SD is an anti-symmetrized product of $N$ single-particle states (called occupied orbitals):\cite{SlaterDeterminant}
\begin{equation}
\ket{\Psi} = \hat{a}_1^{\dagger} \hat{a}_2^{\dagger} ...  \hat{a}_N^{\dagger} \ket{-} \label{theSDeq}
\end{equation}
The variational freedom is a unitary transformation from the given single-particle basis of $L$ orbitals to another basis where the first $N$ orbitals are used to construct the SD. There is gauge freedom in the ansatz, as any unitary transformation that does not mix the $N$ occupied orbitals with the $L-N$ virtual orbitals, does not change the wave-function (except for a global phase). An SD is therefore described by the Grassmann manifold $\mathsf{U}_L / (\mathsf{U}_{N} \times \mathsf{U}_{L-N})$, with $\mathsf{U}_k$ the unitary group of $k \times k$ unitary matrices. This manifold has dimension $2N(L-N)$, and can be parametrized by $N(L-N)$ complex numbers.\cite{PhysRevA.22.2362,*PhysRevA.24.673} This will henceforth be called a complex dimension $N(L-N)$.

\subsection{The Fock equations}
If the particles of a system interact pairwise, the Hamiltonian can always be written in second quantization as
\begin{equation}
\hat{H} = \hat{b}^{\dagger}_i \mathsf{T}^{i}_{~j} \hat{b}^j + \frac{1}{2} \hat{b}^{\dagger}_i \hat{b}^{\dagger}_j \mathsf{V}^{ij}_{~~kl} \hat{b}^l \hat{b}^k. \label{HamiltonianSecondQuantization}
\end{equation}
The TIVP can be expressed in terms of the unitary transformation generating the occupied orbitals:
\begin{widetext}
\begin{equation}
\mathcal{L} = U^{\dagger \alpha}_{\quad i} \mathsf{T}^{i}_{~j} U^{j}_{~\alpha} + \frac{1}{2} U^{\dagger \alpha}_{\quad i} U^{\dagger \beta}_{\quad j} \mathsf{V}^{ij}_{~~kl} U^{l}_{~\beta} U^{k}_{~\alpha} - \frac{1}{2} U^{\dagger \alpha}_{\quad i} U^{\dagger \beta}_{\quad j} \mathsf{V}^{ij}_{~~kl} U^{l}_{~\alpha} U^{k}_{~\beta} - \lambda_{~\alpha}^{\beta} \left(  U^{\dagger \alpha}_{\quad k} U^{k}_{~\beta} - \delta^{\alpha}_{~\beta} \right).
\end{equation}
The Greek indices denote occupied orbitals, while the Latin indices denote all single-particle basis states. Varying with respect to $U^{\dagger m}_{\quad i}$, leads to the Fock equations:\cite{Hartree1,*Hartree2,*SlaterSCF,*FockHF}
\begin{equation}
\mathsf{F}^{i}_{~k} U^{k}_{~p} = \left( \mathsf{T}^{i}_{~k} + U^{\dagger \beta}_{\quad j} \mathsf{V}^{ij}_{~~kl} U^{l}_{~\beta} - U^{\dagger \beta}_{\quad j} \mathsf{V}^{ij}_{~~lk} U^{l}_{~\beta} \right)  U^{k}_{~p} = U^{i}_{~q} \lambda_{~p}^{q}. \label{HFequations}
\end{equation}
\end{widetext}
The gauge can be partially fixed by requiring that the Lagrangian multiplier matrix $\lambda$ to enforce orthonormal orbitals becomes diagonal, and that the diagonal elements are sorted in ascending order. These diagonal elements are then interpreted as the single-particle energy levels.\cite{helgaker2, *helgaker:52, PhysRevA.22.2362,*PhysRevA.24.673} The remaining gauge freedom is then $\mathsf{U}_1^{~\otimes L}$, i.e. the phase of each HF orbital. The lowest $N$ single-particle states are used to construct the SD.\\
The Fock equations are orbital-based mean-field equations. There is self-consistency because the Fock operator in Eq. \eqref{HFequations} which determines the orbitals, also depends on the orbitals.

\subsection{The Thouless theorem \label{linktoToulessHF}}
The Thouless theorem for HF\cite{Thouless1960225,*Thouless196178} and its unitary counterpart\cite{PhysRevA.22.2362,*PhysRevA.24.673} state that any $N$-electron SD can be globally parametrized as resp.
\begin{eqnarray}
\ket{\Psi} & \propto & \exp{\left( X^{vo} \hat{B}_{vo}^{\dagger} \right)} \ket{\Psi^0} \label{HFThouless}\\
\ket{\Psi} & = & \exp{\left( X^{vo} \hat{B}_{vo}^{\dagger} - \overline{X}_{vo} \hat{B}^{vo} \right)} \ket{\Psi^0} \label{counterpartThouless}
\end{eqnarray}
with $\ket{\Psi^0}$ a random SD, with $N$ occupied ($o$) and $L-N$ virtual ($v$) orbitals. $\hat{B}_{vo}^{\dagger}$ is a shorthand for $\hat{a}_{v}^{\dagger} \hat{a}^{o}$. Note that the summation convention was used. The equality in Eq. \eqref{counterpartThouless} holds because the exponential of an anti-hermitian operator is unitary, and hence does not change the norm.\\
This parametrization is of complex dimension $N(L-N)$, and is thus non-redundant. All parameters $X$ are needed to parametrize the neighbourhood of $\ket{\Psi^0}$. For all possible combinations $\ket{\Psi^0}$ and $\ket{\Psi}$, a solution $X$ can always be found. The theorem does not state that this $X$ is unique. In fact, $X$ is not unique, see e.g. the discussion in the appendix of \textcite{PhysRevA.22.2362} The reader can think about the Lie group of $\mathsf{O}(3)$, where several combinations of successive rotations along different axes can generate the same global rotation. Instead of working with the redundant parameters $U$, we can equivalently work with the non-redundant parameters $X$.\\
The $n^{\text{th}}$ order variation of a wave-function defines its $n^{\text{th}}$ order tangent space. The (first order) tangent space of this non-redundant parametrization consists of the single OV excitations $\hat{B}_{vo}^{\dagger} \ket{\Psi^0}$.

\subsection{Time evolution}
The TDVP leads to the time-dependent SCF equations:\cite{PSP:2040328, *RevModPhys.44.602, *Kerman1976332, *Kramer,ringschuck,PhysRevA.22.2362,*PhysRevA.24.673}
\begin{equation}
i \hbar \dot{U}(t)^{i}_{~ p} = F(t)^{i}_{~k} U(t)^{k}_{~ p} \label{TDHF}
\end{equation}
The Fock operator dictates how orbitals are rotated into each other over time. Rotations within the space of occupied orbitals or within the space of virtual orbitals, do not change the SD wavefunction as it represents a Grassmann manifold. Only the rotation of occupied and virtual orbitals into each other has physical meaning.\\
To obtain the rate of OV rotation determined by Eq. \eqref{TDHF} in the point $\ket{\Psi^0}$, the Thouless parametrization of a general SD can be used in the TDVP:
\begin{equation}
i \hbar \dot{X}^{vo}( ~\mathbf{X}=\mathbf{0}~ ) = \braket{\overline{\Psi^0} \mid \hat{B}^{vo} \hat{H} \mid \Psi^0}. \label{OVrotations}
\end{equation}
The parameters $X^{vo}$ are flattened to a column $\mathbf{X}$. The same equation is obtained by inserting Eq. \eqref{counterpartThouless} in the time-dependent Schr\"odinger equation, and by projecting this equation onto $\hat{B}_{vo}^{\dagger} \ket{\Psi} = \hat{B}_{vo}^{\dagger} \ket{\Psi(\mathbf{X},\overline{\mathbf{X}})}$. The time evolution and its projection are resp. given by
\begin{eqnarray}
& & i \hbar \left( \dot{X}^{wp} \frac{\partial}{\partial X^{wp}} + \dot{\overline{X}}_{wp}  \frac{\partial}{\partial \overline{X}_{wp}} \right) \ket{\Psi} = \left( \hat{H} - E_{\text{HF}} \right) \ket{\Psi} ~ \quad\\
& & i \hbar \bra{\overline{\Psi}} \hat{B}^{vo} \left( \dot{X}^{wp} \frac{\partial}{\partial X^{wp}} + \dot{\overline{X}}_{wp}  \frac{\partial}{\partial \overline{X}_{wp}} \right) \ket{\Psi} \nonumber\\
& & \qquad \qquad \qquad \qquad \qquad = \bra{\overline{\Psi}} \hat{B}^{vo} \left( \hat{H} - E_{\text{HF}} \right) \ket{\Psi} \label{toLinearize}
\end{eqnarray}
Evaluation for $\mathbf{X} = \mathbf{0}$ yields Eq. \eqref{OVrotations}.

\subsection{RPA}
Linearization of the TDVP near a variational minimum leads to RPA.\cite{ringschuck, Zyrianov,*PhysRev.107.450,*PhysRev.107.1631, PhysRevA.22.2362,*PhysRevA.24.673} Take the variational minimum as the reference $\ket{\Psi^0}$. Expand Eq. \eqref{toLinearize} up to first order around $\mathbf{X} = \mathbf{0}$. The zeroth order terms vanish because the expansion point is a variational minimum: $\braket{\overline{\Psi^0} \mid \hat{B}^{vo} \hat{H} \mid \Psi^0} = 0$. This is Brillouin's theorem.\cite{Brillouin} The linearized equations are:
\begin{eqnarray}
i \hbar \dot{X}^{vo} & = & \braket{\overline{\Psi^0} \mid \hat{B}^{wp} \hat{B}^{vo} \left(\hat{H} - E_{\text{HF}} \right) \mid \Psi^0} \overline{X}_{wp} \nonumber\\
& + & \braket{\overline{\Psi^0} \mid \hat{B}^{vo} \left(\hat{H} - E_{\text{HF}} \right) \hat{B}^{\dagger}_{wp} \mid \Psi^0} X^{wp}\\
& = & - \braket{\overline{\Psi^0} \mid \left[ \hat{B}^{vo}, \left[ \hat{H} , \hat{B}^{wp} \right] \right] \mid \Psi^0} \overline{X}_{wp} \nonumber\\
& + & \braket{\overline{\Psi^0} \mid \left[ \hat{B}^{vo} , \left[ \hat{H} , \hat{B}^{\dagger}_{wp} \right] \right] \mid \Psi^0} X^{wp}
\end{eqnarray}
Assume a harmonic motion of the form $\mathbf{X} = \mathbf{Y} e^{-i \omega t} + \overline{\mathbf{Z}} e^{i \omega t}$. This leads to the RPA equations:
\begin{equation}
\hbar \omega \left[ \begin{array}{cc} I & 0 \\ 0 & -I \end{array} \right] \left( \begin{array}{c} \mathbf{Y} \\ \mathbf{Z} \end{array} \right) = \left[ \begin{array}{cc} A & B \\ \overline{B} & \overline{A} \end{array} \right] \left( \begin{array}{c} \mathbf{Y} \\ \mathbf{Z} \end{array} \right) \label{eqRPAunit}
\end{equation}
with $A_{vo;wp} = \braket{\overline{\Psi^0} \mid \left[ \hat{B}^{vo} , \left[ \hat{H}, \hat{B}^{\dagger}_{wp} \right] \right] \mid \Psi^0}$ and $B_{vo;wp} = - \braket{\overline{\Psi^0} \mid \left[ \hat{B}^{vo} , \left[ \hat{H}, \hat{B}^{wp} \right] \right] \mid \Psi^0}$. Note that if $(\omega, \mathbf{Y}, \mathbf{Z})$ is a solution, $(-\omega, \overline{\mathbf{Z}}, \overline{\mathbf{Y}})$ is a solution too.\\
Consider the energy functional
\begin{equation}
E(\mathbf{X},\overline{\mathbf{X}}) = \braket{\overline{\Psi(\mathbf{X},\overline{\mathbf{X}})} \mid \hat{H} \mid \Psi(\mathbf{X},\overline{\mathbf{X}})}
\end{equation}
and its expansion up to second order in $\mathbf{X}$:
\begin{equation}
E^{(2)}(\mathbf{X},\overline{\mathbf{X}}) - E_{\text{HF}} = \frac{1}{2} \left( \begin{array}{c} \mathbf{X} \\ \overline{\mathbf{X}} \end{array} \right)^{\dagger} \left[ \begin{array}{cc} A & B \\ \overline{B} & \overline{A} \end{array} \right]  \left( \begin{array}{c} \mathbf{X} \\ \overline{\mathbf{X}} \end{array} \right).
\end{equation}
The RPA method searches for the harmonic modes of this potential near its minimum, akin to normal mode analysis in analytical mechanics.\\
In linear response theory, the RPA frequencies occur as poles in the response function. Because the exact response function for the exact ground state has the excitation energies of the Hamiltonian as poles, the RPA frequencies are interpreted as approximate excitation energies.\cite{ringschuck} A second argument to interpret the RPA frequencies as excitation energies is given by the alternative derivation of RPA by means of the EOM.\cite{ringschuck, PhysRev.106.372, *PhysRev.108.507, *PhysRev.112.1900} Assume we know the exact ground state $\ket{0}$, and the exact excitation operators which connect the ground state to the excited states $\hat{Q}^{\dagger}_n = \ket{n}\bra{\overline{0}}$. The operator $\hat{Q}_n$ then destroys the ground state. With $\hbar \omega_n = E_n - E_0$, the excitation energy of the excited state $\ket{n}$, it is easy to derive the EOM:
\begin{equation}
\braket{ \overline{0} \mid \left[ \delta \hat{Q}, \left[ \hat{H}, \hat{Q}^{\dagger}_n \right] \right] \mid 0 } = \hbar \omega_n \braket{ \overline{0} \mid \left[ \delta \hat{Q}, \hat{Q}^{\dagger}_n \right] \mid 0}. \label{EOMeq}
\end{equation}
For RPA, two assumptions are made: the excitation operators are approximated by $\hat{Q}_n^{\dagger} = Y^{vo}_n \hat{B}_{vo}^{\dagger} - Z^{vo}_n \hat{B}^{vo}$, and the expectation values of the commutators are calculated with the HF reference wavefunction. The latter approximation is called the quasi-boson approximation.  The RPA Eq. \eqref{eqRPAunit} is then retrieved.\\
As an alternative, an exact bosonic algebra can be set up:
\begin{eqnarray}
\left[\hat{B}^{vo},\hat{B}_{wp}^{\dagger}\right] & = & \delta^{v}_{w} \delta^{o}_{p}\\
\left[\hat{B}^{vo},\hat{B}^{wp}\right] & = & 0
\end{eqnarray}
by adding higher order terms to $\hat{B}_{vo}^{\dagger} = \hat{a}^{\dagger}_v \hat{a}_o + \mathcal{O}(\hat{a}^4)$. The Hamiltonian can then be written in terms of these new operators:
\begin{equation}
\hat{H}_B = E_{\text{HF}} - \frac{\text{tr}A}{2} + \frac{1}{2} \left( \hat{\mathbf{B}}^{\dagger} \hat{\mathbf{B}} \right) \left[ \begin{array}{cc} A & B \\ \overline{B} & \overline{A} \end{array} \right] \left(  \begin{array}{c} \hat{\mathbf{B}}\\ \hat{\mathbf{B}}^{\dagger} \end{array} \right) + \mathcal{O}(\hat{B}^3).
\end{equation}
RPA coincides with neglecting all terms of $\mathcal{O}(\hat{B}^3)$ in the bosonic expansion. This leads to the RPA correlation energy and wavefunction:\cite{ringschuck,Thouless1960225,*Thouless196178}
\begin{eqnarray}
E_{\text{cRPA}} & = & - \frac{\text{tr}A}{2} + \sum\limits_{\omega_n>0} \frac{\hbar \omega_n}{2} = - \sum\limits_{n} \hbar \omega_n \sum\limits_{vo} \mid Z_{n}^{vo} \mid^2 \label{RPAcorrEnergyEq}\quad\\
\ket{\text{RPA}} & \propto & e^{\frac{1}{2} \left( \overline{ZY^{-1}} \right)^{vo;wp} \hat{B}^{\dagger}_{vo} \hat{B}^{\dagger}_{wp}} \ket{\Phi^0}. \label{RPAwavefunction}
\end{eqnarray}
The RPA correlation energy has contributions from the zero point energy of the harmonic oscillators with frequency $\omega_n$. The RPA wavefunction vanishes by the action of deexcitation operators: $\hat{Q}_n \ket{\text{RPA}} = 0$.\\
If the Hamiltonian has a continuous symmetry, and the exact ground state is degenerate due to this symmetry, a ground state calculation typically breaks this symmetry. Think for example about a spin-$\frac{1}{2}$ ground state. A calculation will lead to one possibility: $\alpha \ket{s^z = \frac{1}{2}} + \beta \ket{s^z = -\frac{1}{2}}$. A  gapless bosonic degree of freedom remains, which corresponds to rotating within the spin-$\frac{1}{2}$ multiplet, called a Goldstone boson.\cite{PhysRev.117.648,*GoldstoneBoson} An interesting feature of RPA is its ability to retrieve Goldstone modes. The excitation energy of a Goldstone mode is of course zero, and the mode is its own dual solution $(\omega=0, \mathbf{Y}, \mathbf{Z}) = (\omega=0, \overline{\mathbf{Z}}, \overline{\mathbf{Y}})$.\cite{Thouless1960225,*Thouless196178,DimitriBook} This implies that
\begin{equation}
\sum_{vo} \left( \overline{Y}_{vo} Y^{vo} - \overline{Z}_{vo} Z^{vo} \right) = 0. \label{GoldstoneEq}
\end{equation}

\subsection{Post-HF methods}
With the excitation operators $\hat{B}^{\dagger}_{vo} = \hat{a}_{v}^{\dagger} \hat{a}^o$, a set of orthonormal vectors can be generated: $\ket{\text{HF}}$, $\hat{B}^{\dagger}_{vo} \ket{\text{HF}}$, $ \hat{B}^{\dagger}_{vo} \hat{B}^{\dagger}_{wp} \ket{\text{HF}}$... They correspond to the zeroth, first and second order tangent space of the Thouless parametrization of a general SD. With the CI method, eigenstates of the Hamiltonian are approximated by working in an incomplete basis of such vectors.\cite{SlaterDeterminant,PhysRev.36.1121} Consider for example the second order expansion CISD, or CI with single and double excitations:
\begin{equation}
\ket{\text{CISD}} \propto \left( x + y^{vo}\hat{B}^{\dagger}_{vo} + \frac{1}{2} z^{vo;wp} \hat{B}^{\dagger}_{vo} \hat{B}^{\dagger}_{wp} \right) \ket{\text{HF}}
\end{equation}
With CIS, or CI with only single excitations, the lowest energy state is again $\ket{\text{HF}}$ due to Brillouin's theorem, and the eigenstates approximated in the basis $\hat{B}^{\dagger}_{vo} \ket{\text{HF}}$ are therefore excited states. Note that this corresponds to diagonalizing the A-matrix of RPA in Eq. \eqref{eqRPAunit}. Historically, this method is known as TDA.\cite{Tamm, *PhysRev.78.382}\\
The RPA wavefunction in Eq. \eqref{RPAwavefunction} suggests a CC ansatz.\cite{Coester1958421,*Coester1960477,*cizek:4256} Consider for example CCSD, or CC with single and double excitations:
\begin{equation}
\ket{\text{CCSD}} \propto e^{\left( y^{vo}\hat{B}^{\dagger}_{vo} + \frac{1}{2} z^{vo;wp} \hat{B}^{\dagger}_{vo} \hat{B}^{\dagger}_{wp} \right)} \ket{\text{HF}}
\end{equation}
An important property of ansatz wave-functions is their size consistency, i.e. the property that for two non-interacting subsystems, the compound wave-function is multiplicatively separable and the total energy additively separable. CISD is not size consistent if there are more than two electrons in the compound system, whereas CCSD is always size consistent because of the exponential ansatz.\cite{helgaker2,*helgaker:52}

\section{The matrix product state}
\subsection{The ansatz}
Consider the many-body Hilbert space $\ket{n_1 n_2 ... n_L}$, formed by taking the direct product of $L$ local Hilbert spaces $\ket{n_i}$. The local degrees of freedom can be e.g. the spin projections of spins on a lattice, or the occupancies of orbitals. In the latter case, the states $\ket{n_1 n_2 ... n_L}$ form the Fock space.\cite{FockSpace} An MPS can be seen as a linear combination of these vectors, where the coefficient of each vector is a product of matrices:
\begin{equation}
\ket{\Phi} = \sum\limits_{\{ n_i \}} A[1]^{n_1} A[2]^{n_2}  ... A[L]^{n_L} \ket{n_1 n_2 ... n_L}. \label{EqWhatIsAnMPS}
\end{equation}
We assume an MPS with open boundary conditions, i.e. the first matrix has row dimension 1 and the last matrix has column dimension 1. The bond dimension (virtual dimension) $D_i$ of an MPS at boundary $i$ is the column dimension of the matrices at site $i$ and the row dimension of the matrices at site $i+1$. With our assumption, $D_0=D_L=1$. The total number of complex parameters in this ansatz is $\text{dim} \mathbb{A} = \sum_{i=1}^{L} D_{i-1} d D_{i}$, with $d$ the size of the local Hilbert space $\ket{n_i}$. Just as in the SD, there is gauge freedom in the ansatz. Right-multiplying the $d$ site-matrices on site $i$ with the non-singular matrix $G$ ($\tilde{A}[i]^{n_i} = A[i]^{n_i} G$), and simultaneously left-multiplying the $d$ site-matrices on site $i+1$ with its inverse $G^{-1}$ ($\tilde{A}[i+1]^{n_{i+1}} = G^{-1} A[i+1]^{n_{i+1}}$), does not change the wave-function ($\tilde{A}[i]^{n_i} \tilde{A}[i+1]^{n_{i+1}} = A[i]^{n_i} A[i+1]^{n_{i+1}}$). A global scalar multiplication does not change the wave-function either. The MPS manifold, i.e. the quotient space of the general parametrization (complex dimension $\text{dim} \mathbb{A}$) and all gauge freedom (complex dimension $\sum_{i=1}^L D_i^2$), has complex dimension $\text{dim} \mathbb{T} = \sum_{i=1}^{L} \left( dD_{i-1}-D_i \right) D_i$.\cite{PhysRevLett.107.070601,2012arXiv1210.7710H,TTstuff,*Uschmajew2013133}

\subsection{The SD as low bond dimension limit}
An interesting connection to HF can be made by considering an MPS where the $L$ orbitals are the HF orbitals. As each orbital occupation number is definite in an SD, an MPS with bond dimension 1 suffices to represent it. Conversely, if an MPS has bond dimension 1 and represents a state with definite particle number, each orbital has a definite occupation number. If this is not the case, two or more orbitals must be entangled (there is static correlation between them), and the bond dimension has to be larger than 1 to represent this. An MPS with bond dimension 1 and definite particle number can hence always be represented by an SD. An SD is the low bond dimension limit of an MPS, while a general full CI (FCI) solution requires an exponentially large bond dimension to be represented by an MPS.\cite{Schollwoeck}\\
The SD ansatz provides a single variational approximation to the ground state, which unfortunately fails to represent static correlation. On the contrary, the MPS ansatz allows to systematically improve the approximation to the ground state by increasing the bond dimension, up to the point where all static correlation is resolved.\cite{white:4127, *chan:4462, wouters, *sharma:124121, chan:annurevphys}

\section{The DMRG equations}
The TIVP leads to the DMRG equations.\cite{B805292C} The canonical DMRG equations for site $i$ are retrieved when additional constraints are added to the Lagrangian to enforce that the site-matrices to the left of site $i$ are left-normalized:
\begin{equation}
\forall j<i : \sum\limits_{n_j} (A^{n_j}[j])^{\dagger} A^{n_j}[j] = I_{D_j} \label{left-can-index}
\end{equation}
and that the site-matrices to the right of site $i$ are right-normalized:
\begin{equation}
\forall j>i : \sum\limits_{n_j} A^{n_j}[j] (A^{n_j}[j])^{\dagger} = I_{D_{j-1}}.
\end{equation}
With $(A^{n_j}[j])_{\alpha \beta} = A[j]^{n_j \alpha \beta}$, the Lagrangian becomes
\begin{widetext}
\begin{equation}
\mathcal{L} = \braket{\overline{\Phi} \mid \hat{H} \mid \Phi} - \lambda \left( \overline{A[i]_{n_i \alpha \beta}} A[i]^{n_i \alpha \beta}  - 1 \right) - \sum\limits_{j<i} \lambda[j]_{\gamma}^{~\beta} \left( \overline{A[j]_{n_j \alpha \beta}} A[j]^{n_j \alpha \gamma}  - \delta^{~\gamma}_{\beta}  \right) - \sum\limits_{j>i} \lambda[j]_{\gamma}^{~\alpha} \left( \overline{A[j]_{n_j \alpha \beta}} A[j]^{n_j \gamma \beta}  - \delta^{~\gamma}_{\alpha}  \right).
\end{equation}
\end{widetext}
Varying with respect to $\overline{A[i]_{n_i \alpha \beta}}$ gives the canonical one-site DMRG equations:
\begin{equation}
H_{\text{eff}}[i]^{n_i \alpha \beta}_{\quad \tilde{n}_i \tilde{\alpha} \tilde{\beta}} A[i]^{\tilde{n}_i \tilde{\alpha} \tilde{\beta}} = \lambda A[i]^{n_i \alpha \beta} \label{DMRGeffHamEq}
\end{equation}
in terms of the effective Hamiltonian.\cite{B805292C}
By bringing the MPS into canonical forms, of which the left- and right-normalization conditions above are examples, the gauge freedom can be (partially) removed. For the left- and right-normalization conditions, the remaining gauge freedom is a unitary rotation ($G$ unitary). All gauge freedom can be removed by bringing the MPS into Vidal's canonical form.\cite{PhysRevLett.91.147902}\\
The DMRG equations are site-based mean-field equations. There is self-consistency because the effective Hamiltonian in Eq. \eqref{DMRGeffHamEq} which determines the site-matrices of a particular site, depends on the site-matrices of the other sites.\cite{Schollwoeck, B805292C, 2011arXiv1103.2155K,*2013APS..MARJ24006K} In DMRG, the effective Hamiltonian hence plays the role of Fock operator.\cite{B805292C} Since both of them act locally (resp. on one site and one orbital), it might be worthwhile to explore Rayleigh-Schr\"odinger perturbation theory analogues for DMRG in the future, such as M\o{}ller-Plesset perturbation theory.\cite{PhysRev.46.618,B805292C,helgaker2,*helgaker:52}\\
Note that in practice the two-site DMRG algorithm is used to optimize an MPS. The two-site algorithm is more robust against local minima, and when symmetry is imposed it provides a natural way to distribute the bond dimension $D$ over the symmetry sectors. After the two-site algorithm has converged, a few one-site DMRG sweeps allow to make the MPS fully self-consistent. This can be compared to HF, where the optimal SD is found by gradient methods\cite{ringschuck} or by direct inversion of iterative subspaces\cite{Pulay1980393} for stability reasons. The DMRG and HF solutions satisfy resp. Eqs. \eqref{DMRGeffHamEq} and \eqref{HFequations}, irrespective of the optimization scheme.

\section{The MPS tangent space\label{sectiontangentspaces}}
\subsection{A redundant parametrization}
Flatten the site-matrices $A[i]^{n_i}$ to a column $\mathbf{A}$ with entries $\left( A[i]^{n_i} \right)_{\alpha,\beta} = A^{i n_i \alpha \beta} = A^{\mu}$, and consider a small variation $A^{\mu} = A_0^{\mu} + B^{\mu}$. The wave-function can then be expanded as
\begin{equation}
\ket{\Phi} = \ket{\Phi^0} + B^{\mu} \ket{\Phi^0_{\mu}} + \frac{1}{2} B^{\mu} B^{\nu} \ket{\Phi^0_{\mu \nu}} + ... \label{wavefunctionexpansion}
\end{equation}
with first order tangent space $\ket{\Phi^0_{\mu}} = \partial_{\mu} \ket{\Phi^0} = \frac{\partial \ket{\Phi^0}}{\partial A^{\mu}}$ and second order tangent space $\ket{\Phi^0_{\mu \nu}} = \partial_{\mu} \partial_{\nu} \ket{\Phi^0}$. Note that the summation convention was used. Each order of MPS tangent space contains all lower orders: e.g. $A_0^{\mu} \ket{\Phi^0_{\mu}} = L \ket{\Phi^0}$ and $A_0^{\mu} B^{\nu} \ket{\Phi^0_{\mu \nu}} = (L-1) B^{\mu} \ket{\Phi^0_{\mu}} $.\cite{2011arXiv1103.2155K,*2013APS..MARJ24006K, PhysRevLett.107.070601, 2012arXiv1210.7710H}\\
The tangent vectors $\ket{\Phi^0_{\mu}}$ are redundant, as the MPS manifold has dimension $\text{dim} \mathbb{T}$, and there are $\text{dim} \mathbb{A}$ such vectors. The metric, or overlap matrix $S_{\mu\nu} = \braket{\overline{\Phi_{\mu}^0} \mid \Phi_{\nu}^0}$, is therefore not invertible. In Sec. \ref{non-red-section}, $\text{dim} \mathbb{T}$ explicit linear combinations of the vectors $\ket{\Phi^0_{\mu}}$ are given, so that the overlap in this new basis is the unit matrix, and $\ket{\Phi^0}$ is orthogonal to this new basis. Remember that variations in the direction of $\ket{\Phi^0}$ only cause norm or phase changes of the ansatz, but do not change the physical state. This new basis is then a non-redundant parametrization of the MPS tangent space.

\subsection{Hamiltonian sparsity}
The Hamiltonian in Eq. \eqref{HamiltonianSecondQuantization} is sparse, as it consists of a sum of one- and two-particle interactions. When it acts on a certain SD, the result lies in the space spanned by the given SD and its single and double OV excitations. This is immediataly clear by changing the single particle basis in Eq. \eqref{HamiltonianSecondQuantization} from $\hat{b}_k^{\dagger}$ to the SD orbitals $\hat{a}^{\dagger}_j$.\\
A typical lattice Hamiltonian can be considered sparse too, as it consists of a sum of one- and two-site operators. It is sparse in site-space instead of particle-space. Let us focus on the one-dimensional Hubbard model:\cite{HubbarModelCite}
\begin{equation}
\hat{H} = - \sum\limits_{\sigma, i=1}^{L-1} \left( \hat{a}^{\dagger}_{i \sigma} \hat{a}_{i+1 \sigma} + \hat{a}^{\dagger}_{i+1 \sigma} \hat{a}_{i \sigma} \right) + U \sum\limits_{i=1}^{L} \hat{n}_{i \uparrow} \hat{n}_{i \downarrow}. \label{HubbardModel}
\end{equation}
Consider its action on an MPS. Let $\mu_i$ be a shorthand for $(n_i,\alpha,\beta)$, or $\mu$ restricted to site $i$. The Hamiltonian connects the MPS to a part of its double tangent space:
\begin{equation}
\hat{H} \ket{\Phi^0} \propto C^{\mu_i \nu_{i+1}} \ket{\Phi^0_{\mu_i \nu_{i+1}}}.
\end{equation}
It might hence be worthwhile to construct the site-space analogue of the particle Fock space.\cite{DiracSecondQuantization,FockSpace} A new second quantization should be constructed, based on the MPS reference instead of the HF orbitals.

\subsection{A non-redundant parametrization\label{non-red-section}}
A non-redundant parametrization of the MPS tangent space was first presented by \textcite{dorando:184111} in DMRG projector terminology. \textcite{PhysRevLett.107.070601} provided a construction in the language of the MPS wave-function and the corresponding manifold. To present the relationship between the two, here we describe the tangent space construction in projector terms, but by using the explicit MPS representation of the projectors.\\
Consider an MPS where all left renormalized basis states at boundary $i-1$:
\begin{equation}
\ket{L_{\alpha}^{i-1}} = \sum\limits_{\{n_j:j<i\}} \left[ A^{n_1}[1] .. A^{n_{i-1}}[i-1] \right]_{\alpha} \ket{n_1 .. n_{i-1}}
\end{equation}
are orthonormal and all right renormalized basis states at boundary $i$:
\begin{equation}
\ket{R_{\beta}^{i}} = \sum\limits_{\{n_j:j>i\}} \left[ A^{n_{i+1}}[i+1] .. A^{n_L}[L] \right]_{\beta} \ket{n_{i+1} .. n_L}
\end{equation}
are orthonormal. In the DMRG algorithm, a renormalization transformation is constructed to reduce the direct product of $\ket{L_{\alpha}^{i-1}}$ (size $D_{i-1}$) and $\ket{n_i}$ (size $d$) to a new left renormalized basis at boundary $i$ (size $D_i \leq dD_{i-1}$). This renormalization transformation is a projection, represented by the site-matrices of site $i$:
\begin{equation}
\sum\limits_{\alpha n_i} \left( A[i]^{n_i} \right)_{\alpha ,\beta} \ket{L_{\alpha}^{i-1}} \ket{n_i}.
\end{equation}
The projection onto the $dD_{i-1}-D_i$ discarded states from the direct product space, defines the non-redundant tangent space. We now explain the explicit construction of the non-redundant tangent space as provided by \textcite{dorando:184111} in MPS terminology. Consider the QR-decomposition of the projector:
\begin{equation}
\left( A[i]^{n_i} \right)_{\alpha, \beta} = A[i]_{(\alpha n_i),\beta} = \sum\limits_{\gamma} Q[i]_{(\alpha n_i),\gamma} R_{\gamma,\beta}. \label{QRdecompeq}
\end{equation}
$Q[i]$ contains $D_i$ orthonormal columns of size $d D_{i-1}$. Its left nullspace is spanned by $d D_{i-1} - D_i$ vectors. This allows to construct the $dD_{i-1} \times (d D_{i-1} - D_i)$ matrix $\tilde{Q}[i]$, so that $\left[ Q[i] \tilde{Q}[i] \right]$ is unitary. A part of the non-redundant tangent space can then be parametrized by the matrix $x[i]$ with dimensions $\left( dD_{i-1}-D_i \right) \times D_i$:
\begin{equation}
\sum\limits_{\alpha \beta n_i} \left( \tilde{Q}[i]^{n_i} x[i] \right)_{\alpha,\beta} \ket{L_{\alpha}^{i-1}} \ket{n_i} \ket{R_{\beta}^{i}}.
\end{equation}
If the renormalized basis states $\ket{L_{\alpha}^{i-1}}$, $\ket{R_{\beta}^{i}}$ are not orthonormal, their overlap has to be taken into account. It was \textcite{PhysRevLett.107.070601} who first presented the parametrization in that case:
\begin{equation}
\sum\limits_{\alpha \beta n_i} \left( l[i-1]^{-\frac{1}{2}} \tilde{Q}[i]^{n_i} x[i] r[i]^{-\frac{1}{2}} \right)_{\alpha, \beta} \ket{L_{\alpha}^{i-1}} \ket{n_i} \ket{R_{\beta}^{i}} \label{JuthoTangentSpace}
\end{equation}
with $l[i-1]$ the density matrix of the left renormalized states $\ket{L_{\alpha}^{i-1}}$ and $r[i]$ the density matrix of the right renormalized states $\ket{R_{\beta}^{i}}$. The QR-decomposition of Eq. \eqref{QRdecompeq} is now performed on $l[i-1]^{\frac{1}{2}} A[i]^{n_i}$ instead of on $A[i]^{n_i}$. The complete non-redundant tangent space is formed by doing this construction for the projector on each site. Combine all matrices $x[i]$ to a column $\mathbf{x}$ of length $\text{dim} \mathbb{T}$. By writing the construction in Eq. \eqref{JuthoTangentSpace} as $B^{\mu}(\mathbf{x}) \ket{\Phi^0_{\mu}}$, with
\begin{equation}
B^{n_i}(\mathbf{x})[i] = l[i-1]^{-\frac{1}{2}} \tilde{Q}^{n_i}[i] x[i] r[i]^{-\frac{1}{2}}, \label{BmxExplicit}
\end{equation}
one possibility for a non-redundant tangent space basis of dimension $\text{dim} \mathbb{T}$ is immediately obtained:
\begin{equation}
\ket{\Phi_k^T} = \frac{\partial}{\partial x^k} B^{\mu}(\mathbf{x}) \ket{\Phi^0_{\mu}}.
\end{equation}
Note that this provides a construction of $\ket{\Phi^T_k}$ as a linear combination of $\ket{\Phi^0_{\mu}}$. Any tangent vector can be constructed by taking a complex linear combination of these $\text{dim} \mathbb{T}$ vectors: $x^k \ket{\Phi^T_k} = B^{\mu}(\mathbf{x}) \ket{\Phi^0_{\mu}}$. Because of the construction of $\tilde{Q}^{n_i}[i]$, these vectors are orthogonal to $\ket{\Phi^0}$: $\braket{\overline{\Phi^0} \mid \Phi^0_{\mu}  } B^{\mu}(\mathbf{x}) = 0$. The metric of the parametrization in Eq. \eqref{BmxExplicit} is the unit matrix: $\overline{B^{\mu}(\mathbf{x})} S_{\mu \nu} B^{\nu}(\mathbf{y}) = \mathbf{x}^{\dagger}\mathbf{y}$.\cite{PhysRevLett.107.070601,2012arXiv1210.7710H} Analogous results have been obtained in a different context.\cite{TTstuff,*Uschmajew2013133}\\
For an SD written as an MPS ($D=1$ and $d=2$), the non-redundant tangent space vectors correspond to the addition (removal) of an electron to (from) the system.

\section{The Thouless theorem for MPS}
The operators $\hat{B}^{\dagger}_{vo}$ link an SD $\ket{\Psi^0}$ to its non-redundant tangent space $\hat{B}^{\dagger}_{vo} \ket{\Psi^0}$. Exponentiation of these operators led to the Thouless theorem. Here we present the MPS counterpart.

\subsection{Proposal}
For the sake of simplicity, we use a part of the gauge freedom to work with a left-canonical MPS. The left-normalization condition in Eq. \eqref{left-can-index} then holds for all sites. This implies $\forall i: l[i] = I_{D_i}$. Consider the following matrix notation for site-matrices: $C[i]$ has entries $(C[i])_{(\alpha n_i),\beta}$, i.e. the row-index of the matrices $C[i]$ contains the physical index $n_i$. The left-normalization condition then becomes
\begin{equation}
A[i]^{\dagger} A[i] = I_{D_i}. \label{left-can-eq}
\end{equation}
Because of the construction of $\tilde{Q}[i]$, the site-matrices $B(\mathbf{x})[i]$ are left-orthogonal to the site-matrices $A[i]$:
\begin{equation}
B(\mathbf{x})[i]^{\dagger} A[i] = 0. \label{B-left-can-eq}
\end{equation}
This allows to propose the MPS counterpart of the Thouless theorem:
\begin{equation}
A(\mathbf{x},\overline{\mathbf{x}})[i] = \exp{\left( B(\mathbf{x})[i] A_0[i]^{\dagger} - A_0[i] B(\mathbf{x})[i]^{\dagger} \right)} A_0[i] \label{ThoulessMPS}
\end{equation}
where now $A_0[i]$ is used for $A[i]$ to clearly mark the difference with $A(\mathbf{x},\overline{\mathbf{x}})[i]$. The matrix in the exponential is anti-Hermitian, and the transformation in Eq. \eqref{ThoulessMPS} is therefore unitary. As the $A_0[i]$ site-matrices were left-normalized, the $A(\mathbf{x},\overline{\mathbf{x}})[i]$ site-matrices are also left-normalized. An MPS built with the $A(\mathbf{x},\overline{\mathbf{x}})[i]$ site-matrices:
\begin{equation}
\ket{\Phi(\mathbf{x},\overline{\mathbf{x}})} = \sum\limits_{\left\{ n_i \right\}} A(\mathbf{x},\overline{\mathbf{x}})[1]^{n_1} ... A(\mathbf{x},\overline{\mathbf{x}})[L]^{n_L} \ket{n_1 n_2 ... n_L} \label{Phi_x_thouless_to_mps}
\end{equation}
is hence still left-canonical and therefore normalized. For $\mathbf{x}=\mathbf{0}$, $\ket{\Phi(\mathbf{x},\overline{\mathbf{x}})} = \ket{\Phi^0}$. The tangent space of this MPS parametrization is familiar too:
\begin{equation}
\left. \frac{\partial}{\partial x^k} \ket{\Phi(\mathbf{x},\overline{\mathbf{x}})} \right|_{\mathbf{x}=\mathbf{0}} = \ket{\Phi^T_k}
\end{equation}
which can be easily checked by using Eqs. \eqref{left-can-eq} and \eqref{B-left-can-eq}. $\ket{\Phi(\mathbf{x},\overline{\mathbf{x}})}$ is therefore an explicit non-redundant parametrization of the MPS manifold in the neighbourhood of $\ket{\Phi^0}$.

\subsection{Global validity}
Here we show that Eq. \eqref{Phi_x_thouless_to_mps} is a global parametrization of the MPS manifold, or that any MPS with bond dimensions $D_i$ can be generated from $\ket{\Phi^0}$ (which has the same bond dimensions). This implies that we can optimize over the parameters $\mathbf{x}$ instead of over $\mathbf{A}$ to find an energy minimum.\\
For a specific site index $i$, the parametrization $A(\mathbf{x},\overline{\mathbf{x}})[i]$ of Eq. \eqref{ThoulessMPS} is a Grassmann manifold with matrix dimensions $dD_{i-1} \times D_i$. Define $\mathbf{y}$ by $x[i] = y[i]r[i]^{\frac{1}{2}}$ to obtain
\begin{eqnarray}
\tilde{A}(\mathbf{y},\overline{\mathbf{y}}) & = & A(\mathbf{x},\overline{\mathbf{x}})[i] \nonumber \\
& = & \exp{\left( \tilde{Q}[i] y Q[i]^{\dagger} - Q[i] y^{\dagger} \tilde{Q}[i]^{\dagger} \right)} Q[i]. \label{GrassmannFormThoulessMPS}
\end{eqnarray}
Note the close analogy to Eq. \eqref{counterpartThouless}. We show in the Appendix that Eq. \eqref{GrassmannFormThoulessMPS} represents a Grassmann manifold. Note that we assume that the density matrix $r[i]$ is non-singular (i.e. $r[i]^{-\frac{1}{2}}$ exists) for the construction of the non-redundant tangent space in Eq. \eqref{BmxExplicit}. For a left-canonical MPS, the Schmidt values are the positive square roots of the eigenvalues of $r[i]$. The condition of non-singular density matrices $r[i]$ is therefore equal to having all Schmidt values of $\ket{\Phi^0}$ non-zero.  This is a condition for the global validity of Thouless's theorem for MPS.\\
Now give a normalized MPS $\ket{\Phi^1}$, with the only restriction that it has the same bond dimensions as $\ket{\Phi^0}$. We will prove by construction
\begin{equation}
\exists \mathbf{x} : \left|\braket{\overline{\Phi^1} \mid \Phi(\mathbf{x},\overline{\mathbf{x}})}\right| = 1.
\end{equation}
\begin{enumerate}
\item Set $i = 1$.
\item Use a part of the gauge freedom at boundary $i$ to bring the site-matrices $A_1[i]$ of $\ket{\Phi^1}$ in left-normalized form: $A^L_1[i]$.
\item Find $x_1[i]$ so that the columns of $A(\mathbf{x_1},\overline{\mathbf{x_1}})[i]$ and the columns of $A^L_1[i]$ span the same space, which is always possible because $A(\mathbf{x},\overline{\mathbf{x}})[i]$ is a Grassmann manifold.
\item Use the remaining gauge freedom at boundary $i$, i.e. a unitary transformation $U_{D_i}$, to enforce $A(\mathbf{x_1},\overline{\mathbf{x_1}})[i] = A^{\text{exact}}_1[i] = A^L_1[i] U_{D_i}$.
\item If $i<L$, set $i = i+1$ and go to 2.
\end{enumerate}
When the construction is finished, all parameters of $\mathbf{x_1}$ are assigned, and the gauge freedom in $\ket{\Phi^1}$ was used to write $\ket{\Phi^1}$ exactly as $\ket{\Phi(\mathbf{x_1},\overline{\mathbf{x_1}})}$, i.e. $\forall i: A^{\text{exact}}_1[i] = A(\mathbf{x_1},\overline{\mathbf{x_1}})[i]$. See Ref. \onlinecite{TTstuff,*Uschmajew2013133} on the diffeomorphism between a finite chain MPS manifold and a product manifold of Grassmann manifolds. This concludes the proof that Eq. \eqref{Phi_x_thouless_to_mps} can represent any MPS with the same bond dimensions, as long as $\ket{\Phi^0}$ does not have any vanishing Schmidt values. Note that the theorem guarantees a solution $\mathbf{x_1}$, but does not guarantee that this solution is unique, in analogy with the discussion in Sec. \ref{linktoToulessHF}. 

\subsection{The double tangent space \label{ThoulessMPSdoubleTangentSpaceSec}}
To get a better understanding of the MPS double tangent space, consider the second order term of $A(\mathbf{x},\overline{\mathbf{x}})[i]$:
\begin{eqnarray}
& & A(\mathbf{x},\overline{\mathbf{x}})[i] - A_0[i] - B(\mathbf{x})[i] \nonumber\\
& = & - \frac{1}{2} A_0[i] B(\mathbf{x})[i]^{\dagger}B(\mathbf{x})[i] + \mathcal{O}(x^3) \nonumber\\
& = & - \frac{1}{2} A_0[i] r[i]^{-\frac{1}{2}} x[i]^{\dagger} x[i] r[i]^{-\frac{1}{2}} + \mathcal{O}(x^3)
\end{eqnarray}
The expansion of $\ket{\Phi(\mathbf{x},\overline{\mathbf{x}})}$ up to second order then consists of
\begin{enumerate}
\item The MPS reference $\ket{\Phi(\mathbf{0},\mathbf{0})} = \ket{\Phi^0}$.
\item The tangent space $\left. \frac{\partial}{\partial x^k} \ket{\Phi(\mathbf{x},\overline{\mathbf{x}})} \right|_{\mathbf{x}=\mathbf{0}} = \frac{\partial}{\partial x^k} \ket{\Phi(\mathbf{0},\mathbf{0})} = \ket{\Phi^T_k}$. Note that $\frac{\partial}{\partial \overline{x_k}} \ket{\Phi(\mathbf{0},\mathbf{0})} = 0$. The tangent space consists of all possible connections between the unused basis states from $\ket{L_{\alpha}^{i-1}} \otimes \ket{n_i}$ and $\ket{R_{\beta}^{i}}$.
\item The non-local part of the double tangent space $\frac{\partial^2}{\partial x^k \partial x^l} \ket{\Phi(\mathbf{0},\mathbf{0})} = \ket{\Phi^{T2}_{kl}}$. Note that this term is only nonzero if $x^k$ and $x^l$ correspond to different sites of the MPS chain. This part corresponds to two single excitations on different sites. Also note that $\frac{\partial^2}{\partial \overline{x_k} \partial \overline{x_l}} \ket{\Phi(\mathbf{0},\mathbf{0})} = 0$.
\item The local part of the double tangent space $\frac{\partial^2}{\partial \overline{x_k} \partial x^l} \ket{\Phi(\mathbf{0},\mathbf{0})} = \ket{\Phi^{T2}_{\overline{k}l}}$. Note that this term is only nonzero if $\overline{x_k}$ and $x^l$ belong to the same site, and correspond to the same row index in the matrix notation $x[i]$. This part consists of all possible connections between the renormalized basis states $\ket{L_{\alpha}^{i}}$ (from $\ket{L_{\alpha}^{i-1}} \otimes \ket{n_i}$) and $\ket{R_{\beta}^{i}}$.
\end{enumerate}
These states are not all mutually orthogonal. Note that the local part of the double tangent space arises because we have considered a unitary variant of the Thouless theorem for MPS. The original (non-unitary) Thouless parametrization for HF depends only on the complex parameters, and not on their complex conjugates.\\
If two excitation operators in HF try to annihilate an occupied single particle twice, the state is destroyed. The space of double OV excitations therefore consists of the replacement of two different occupied single particles by two different virtual single particles.\\
The local part of the double tangent space of an MPS can be written as $B^{\mu} \ket{\Phi^0_{\mu}}$, which lies entirely in the space spanned by the MPS reference $\ket{\Phi^0}$ and the non-redundant tangent space vectors $\ket{\Phi^T_k}$. Together with the other two arguments above, this provides a justification to discard this part of the double tangent space without any loss in variational freedom, and to consider only two single excitations acting on different sites, for the double tangent space.

\subsection{Excitation operators \label{sectionExcOp}}
The excitation operators for an MPS can be read from the Thouless theorem:
\begin{equation}
\ket{\Phi^T_k} = \hat{B}^{\dagger}_k \ket{\Phi^0} =\left. \frac{\partial}{\partial x^k} \ket{\Phi(\mathbf{x},\overline{\mathbf{x}})} \right|_{\mathbf{x}=\mathbf{0}}. \label{ExcitationOperatorEq}
\end{equation}
See e.g. Sec. IV in \textcite{PhysRevA.22.2362} for a discussion on the relationship between the linearized time-dependent variational principle on a general manifold, and the EOM approach to the RPA equations. The operators $\hat{B}^{\dagger}_k$ are obtained by going from the manifold representation based on the virtual space in Eq. \eqref{ThoulessMPS}, to a representation based on the physical Hilbert space $\ket{\Phi(\mathbf{x},\overline{\mathbf{x}})} = \exp{\left( x^k \hat{B}_k^{\dagger} - \overline{x_k} \hat{B}^k \right)} \ket{\Phi^0}$. When only the first order tangent space needs to match, $\hat{B}_k^{\dagger} =  \ket{\Phi^T_k}\bra{\overline{\Phi^0}}$ can be used. It will be a challenge to find the $\hat{B}_k^{\dagger}$'s to match the higher order tangent spaces too. Finding an answer to this problem, is closely related to finding a site-space analogue of the particle Fock space, based on the MPS reference.\\
From Eq. \eqref{ThoulessMPS}, it can be understood that this excitation operator projects out the site-matrices $A_0[i(k)]$ and replaces them with the tangent space site-matrices $\left. \partial_{x_k} B(\mathbf{x})[i(k)] \right|_{\mathbf{x}=\mathbf{0}}$. It adds a single excitation to the vacuum $\ket{\Phi^0}$. In the chosen gauge, a single MPS excitation is localized to one site, just like a single OV excitation of an SD is localized to one orbital.\\
From Eq. \eqref{ThoulessMPS}, it can also be understood that a deexcitation operator projects out the tangent space site-matrices $\left. \partial_{x_k} B(\mathbf{x})[i(k)] \right|_{\mathbf{x}=\mathbf{0}}$ and replaces them with the site-matrices $A_0[i(k)]$. Remember that the tangent space metric is the unit matrix for the chosen parametrization, and that the deexcitation projections are hence not only orthogonal to the MPS reference (they destroy the vacuum $\ket{\Phi^0}$), but also orthogonal to other tangent space site-matrices:
\begin{eqnarray}
\hat{B}_l \ket{\Phi^0} & = & 0\\
\hat{B}_l \ket{\Phi^T_k} & = & \delta_{l,k} \ket{\Phi^0}.
\end{eqnarray}
The deexcitation operators of the ket vectors, are the excitation operators of the bra vectors:
\begin{equation}
\bra{\overline{\Phi^0}} \hat{B}_l = \bra{\Phi^T_l}
\end{equation}
Consider the commutators $\left[ \hat{B}^{\dagger}_l, \hat{B}^{\dagger}_k \right]$ and $\left[ \hat{B}_l, \hat{B}^{\dagger}_k \right]$. Their general expressions are far from trivial, only their expectation value with respect to the vacuum $\ket{\Phi^0}$ is clear:
\begin{eqnarray}
\braket{\overline{\Phi^0} \mid \left[ \hat{B}^{\dagger}_l, \hat{B}^{\dagger}_k \right] \mid \Phi^0 } & = & 0 \label{quasi-boson1}\\
\braket{\overline{\Phi^0} \mid \left[ \hat{B}_l, \hat{B}^{\dagger}_k \right] \mid \Phi^0 } & = & \delta_{k,l}. \label{quasi-boson2}
\end{eqnarray}
A bosonic algebra for the excitation operators is hence only retrieved when expectation values with respect to the vacuum are taken. The operators $\hat{B}^{\dagger}_k$ are called quasi-boson operators.

\section{Optimal time evolution for MPS \label{linktoTimeEvoMPS}}
The optimal time evolution of an MPS, which stays within the MPS ansatz space, was derived by means of the TDVP in Ref. \onlinecite{PhysRevLett.107.070601} and \onlinecite{2011arXiv1103.2155K, *2013APS..MARJ24006K}. Now that we have established the Thouless theorem for MPS, we can rephrase the result as
\begin{equation}
i \hbar \dot{x}^k (\mathbf{x}=\mathbf{0}) = \braket{\overline{\Phi^0} \mid \hat{B}^k \hat{H} \mid \Phi^0}. \label{TDVPquantumLattices}
\end{equation}
Also in this case, Eq. \eqref{TDVPquantumLattices} can be obtained by inserting $\ket{\Phi(\mathbf{x},\overline{\mathbf{x}})}$ in the time-dependent Schr\"odinger equation, and by projecting the time-dependent equation onto $\hat{B}^{\dagger}_k \ket{\Phi(\mathbf{x},\overline{\mathbf{x}})}$:
\begin{equation}
i \hbar \bra{\overline{\Phi}} \hat{B}^{k} \left( \dot{x}^{l} \frac{\partial}{\partial x^l} + \dot{\overline{x}}_{l} \frac{\partial}{\partial \overline{x_l}} \right) \ket{\Phi} = \bra{\overline{\Phi}} \hat{B}^{k} \left( \hat{H} - E_{\text{MPS}} \right) \ket{\Phi} \label{toLinearize2}
\end{equation}
Evaluation for $\mathbf{x}=\mathbf{0}$ yields Eq. \eqref{TDVPquantumLattices}. This form of time propagation automatically stays within the MPS ansatz space. No Hamiltonian decompositions or bond dimension truncations are necessary.\cite{2011arXiv1103.2155K,*2013APS..MARJ24006K,PhysRevLett.107.070601}

\section{RPA for MPS \label{RPAsectionMPS}}
\subsection{In a redundant parametrization}
One way to obtain the RPA equations for MPS, is to consider the linearized time-dependent equations in the vicinity of a variational minimum, and to project them onto the tangent space of the manifold.\cite{2011arXiv1103.2155K,*2013APS..MARJ24006K,juthothesis} Consider a small time-dependent step around the minimum $A^{\mu}(t) = A_0^{\mu} + B^{\mu}(t)$. The time-dependent equation, its projection onto the tangent space, and its first order terms become:
\begin{equation}
i \hbar \ket{\Phi_{\nu}(\mathbf{A})} \dot{A}^{\nu} = \left( \hat{H} - E_{\text{MPS}} \right) \ket{\Phi(\mathbf{A})}
\end{equation}
\begin{widetext}
\begin{eqnarray}
i \hbar \braket{\overline{\Phi_{\mu}(\mathbf{A})} \mid \Phi_{\nu}(\mathbf{A})} \dot{A}^{\nu} & = & \braket{\overline{\Phi_{\mu}(\mathbf{A})} \mid \hat{H} - E_{\text{MPS}} \mid \Phi(\mathbf{A})}\\
i \hbar \braket{\overline{\Phi_{\mu}(\mathbf{A}_0)} \mid \Phi_{\nu}(\mathbf{A}_0)} \dot{B}^{\nu} & = & \braket{\overline{\Phi_{\mu}(\mathbf{A}_0)} \mid \hat{H} - E_{\text{MPS}} \mid \Phi_{\nu}(\mathbf{A}_0)} B^{\nu} + \braket{\overline{\Phi_{\mu\nu}(\mathbf{A}_0)} \mid \hat{H} - E_{\text{MPS}} \mid \Phi(\mathbf{A}_0) } \overline{B}^{\nu}.
\end{eqnarray}
\end{widetext}
with $E_{\text{MPS}} = \braket{\overline{\Phi^0} \mid \hat{H} \mid \Phi^0}$. By taking a harmonic ansatz for the perturbation $\mathbf{B} = \mathbf{Y} e^{-i \omega t} + \overline{\mathbf{Z}} e^{i \omega t}$, the RPA equations are found:
\begin{equation}
\hbar \omega \left[ \begin{array}{cc} S & 0 \\ 0 & -\overline{S} \end{array} \right] \left( \begin{array}{c} \mathbf{Y} \\ \mathbf{Z} \end{array} \right) = \left[ \begin{array}{cc} H & W \\ \overline{W} & \overline{H} \end{array} \right] \left( \begin{array}{c} \mathbf{Y} \\ \mathbf{Z} \end{array} \right) \label{RPAeqWithOverlap}
\end{equation}
with $S_{\mu\nu} = \braket{\overline{\Phi_{\mu}^0} \mid \Phi_{\nu}^0}$, $H_{\mu\nu} = \braket{\overline{\Phi_{\mu}^0} \mid \hat{H} - E_{\text{MPS}} \mid \Phi_{\nu}^0}$, and $W_{\mu\nu} = \braket{\overline{\Phi_{\mu\nu}^0} \mid \hat{H} - E_{\text{MPS}} \mid \Phi^0}$. Note that $\| \left( \hat{H} - E_{\text{MPS}} \right) \ket{\Phi^0} \|_2$ becomes smaller when $\ket{\Phi^0}$ becomes a better approximation for the exact ground state. $\| W \|_2$ is hence a measure for the accuracy of the MPS approximation to the exact ground state.\cite{juthothesis} A specific eigenvector of Eq. \eqref{RPAeqWithOverlap} can be obtained in $\mathcal{O}(L D^3)$ complexity.\cite{dorando:184111,2011arXiv1103.2155K,*2013APS..MARJ24006K}

\subsection{In a non-redundant parametrization}
By changing the basis from $\ket{\Phi_{\mu}^0}$ to $\ket{\Phi^T_k} = Z^{\mu}_k \ket{\Phi_{\mu}^0}$, with $\frac{\partial}{\partial x^k} B^{\mu}(\mathbf{x}) = Z^{\mu}_k$, the overlap matrix $S$ becomes the unit matrix $I$: $\overline{Z}^{\mu}_k S_{\mu \nu} Z^{\nu}_l = \delta_{kl}$. Analogously, the hermitian matrix $A$ and the complex symmetric matrix B, both of dimension $\text{dim} \mathbb{T} \times \text{dim} \mathbb{T}$, can be defined as resp. $A_{kl} = \overline{Z}^{\mu}_k H_{\mu \nu} Z^{\nu}_l$ and $B_{kl} = \overline{Z}^{\mu}_k W_{\mu \nu} Z^{\nu}_l$. The RPA equations become:
\begin{equation}
\hbar \omega \left[ \begin{array}{cc} I & 0 \\ 0 & -I \end{array} \right] \left( \begin{array}{c} \mathbf{y} \\ \mathbf{z} \end{array} \right) = \left[ \begin{array}{cc} A & B \\ \overline{B} & \overline{A} \end{array} \right] \left( \begin{array}{c} \mathbf{y} \\ \mathbf{z} \end{array} \right) \label{RPAeqUNITMPS}
\end{equation}
where $\mathbf{y}$ and $\mathbf{z}$ are now coefficients with respect to $\ket{\Phi^T_k}$. The same result can be obtained by linearizing Eq. \eqref{toLinearize2}, just as for HF.\\
The $A$- and $B$-matrices can be constructed explicitly. If only a few excitation energies are desired, it is better to resort to a sweep algorithm, which can be implemented in an existing DMRG code.\cite{dorando:184111} Implementation details of this sweep algorithm will be presented elsewhere.

\subsection{EOM derivation \label{EOMderivationRPA}}
The excitation operators discussed in Sec. \ref{sectionExcOp} allow for a rederivation of the RPA equations for MPS by means of the EOM. An exact bosonic algebra can be set up by adding correction terms to operators defined in Sec. \ref{sectionExcOp}, so that $\left[ \hat{B}^{\dagger}_l, \hat{B}^{\dagger}_k \right] = 0$ and $\left[ \hat{B}^l, \hat{B}^{\dagger}_k \right] = \delta_{l,k}$. A justification is given by Eqs. \eqref{quasi-boson1} and \eqref{quasi-boson2}. The Hamiltonian can be expanded in these bosonic operators, and RPA coincides with truncating the expansion after second order. Expressions for the RPA correlation energy and wave-function follow, just as for HF:
\begin{eqnarray}
E_{\text{cRPA}} & = & - \frac{\text{tr}A}{2} + \sum\limits_{\omega_n>0} \frac{\hbar \omega_n}{2} = - \sum\limits_{n} \hbar \omega_n \sum\limits_{k} \mid z_{n}^{k} \mid^2 \label{RPAcorrEnergyEq2}\quad\\
\ket{\text{RPA}} & \propto & e^{\frac{1}{2} \left( \overline{zy^{-1}} \right)^{k;l} \hat{B}^{\dagger}_{k} \hat{B}^{\dagger}_{l}} \ket{\Phi^0}. \label{RPAwavefunction2}
\end{eqnarray}

\section{Post-DMRG methods}
\subsection{TDA and Brillouin's theorem}
A preferred tangent basis can be found by searching the eigenstates of the Hamiltonian in the basis $\ket{\Phi^T_k}$. This corresponds to diagonalizing the A-matrix of the RPA Eq. \eqref{RPAeqUNITMPS}. As $\ket{\Phi^0} = \frac{1}{L} A^{\mu}_0 \ket{\Phi^0_{\mu}}$ gave the lowest energy solution for the ansatz $B^{\mu} \ket{\Phi^0_{\mu}}$ and $\ket{\Phi^T_k} \perp \ket{\Phi^0}$, approximations for excited states are found this way. This is the MPS analogue of TDA.\cite{helgaker2, *helgaker:52, Tamm, *PhysRev.78.382, ringschuck} For a variational minimum
\begin{equation}
0 = \frac{\partial E}{ \partial \overline{A^{\mu}_0}} = \frac{\braket{\overline{\Phi_{\mu}^0} \mid \hat{H} \mid \Phi^0}}{\braket{\overline{\Phi^0} \mid \Phi^0}} - \frac{\braket{\overline{\Phi^0} \mid \hat{H} \mid \Phi^0}}{\braket{\overline{\Phi^0} \mid \Phi^0}^2} \braket{\overline{\Phi_{\mu}^0} \mid \Phi^0}.
\end{equation}
If the wave-function $\ket{\Phi^0}$ is normalized and only norm- and phase-conserving changes $\hat{B}_k^{\dagger} \ket{\Phi^0} \perp \ket{\Phi^0}$ are considered,\cite{dorando:184111,PhysRevLett.107.070601}
\begin{equation}
\braket{\overline{\Phi^T_k} \mid \hat{H} \mid \Phi^0} = \braket{\overline{\Phi^0} \mid \hat{B}^k \hat{H} \mid \Phi^0} = 0.\label{BrillouinEq}
\end{equation}
This is the MPS analogue of Brillouin's theorem.\cite{Brillouin,helgaker2, *helgaker:52} For MPS, excited momentum eigenstates of translationally invariant systems have previously been approximated in the non-redundant tangent space basis.\cite{PhysRevB.85.035130,*PhysRevB.85.100408}

\subsection{CC and CI}
The Thouless theorem for MPS and Eq. \eqref{RPAwavefunction2} suggest CC and CI ansatzes on top of an MPS reference. Consider for example the single and double excitations:
\begin{eqnarray}
\ket{\text{CCSD}} & \propto & e^{y^k \hat{B}^{\dagger}_k + \frac{1}{2} z^{kl} \hat{B}^{\dagger}_k \hat{B}^{\dagger}_l} \ket{\Phi^0}\\
\ket{\text{CISD}} & \propto & \left( x + y^k \hat{B}^{\dagger}_k + \frac{1}{2} z^{kl} \hat{B}^{\dagger}_k \hat{B}^{\dagger}_l \right) \ket{\Phi^0}.
\end{eqnarray}
With the exposition in Sec. \ref{sectiontangentspaces}, \ref{ThoulessMPSdoubleTangentSpaceSec} and \ref{sectionExcOp}, we can also propose the following CCSD and CISD ansatzes:
\begin{eqnarray}
\ket{\text{CCSD}} & \propto & e^{ C^{\mu\nu} \partial_{\mu} \partial_{\nu} } \ket{\Phi^0}\\
\ket{\text{CISD}} & \propto & C^{\mu\nu} \ket{\Phi^0_{\mu\nu}} \label{CISDansatz}
\end{eqnarray}
with $\mathcal{O}(\frac{1}{2} (L d D^2)^2)$ parameters in the symmetric C-tensor. Note that working in the $L^\text{th}$ order tangent space corresponds to the FCI ansatz. For DMRG (HF) the CCSD and CISD ansatzes can be considered as a way to improve the correlation between two sites (electrons) embedded in an approximate environment given by the zeroth order MPS (SD). Since the double tangent space can connect sites that are far apart, this enables the CCSD and CISD expressions to directly build in long-range entanglement. If the zeroth order description fails (static correlation for HF, critical system for DMRG), these ansatzes will fail too. Also for MPS, CISD is not size-consistent if there are more than two sites in the compound system, whereas CCSD is always size-consistent because of the exponential ansatz.

\section{Symmetry-adapted calculations\label{SyAdaptSec}}
For large calculations, symmetry-adapted MPS ansatzes are often used. They allow to search for eigenstates within a specific symmetry sector of the total Hilbert space, and lead to computational advantages in memory and time. An MPS ansatz without symmetry-adaptation can yield an approximate eigenstate which breaks the symmetry. Its tangent space then also contains symmetry-broken vectors. RPA-MPS breaks down if a symmetry multiplet of a non-Abelian group is incomplete at a certain MPS boundary. Therefore we use symmetry-adapted MPS ansatzes for the applications.

\subsection{Tangent space without symmetry adaptation}
First consider an MPS ansatz without symmetry adaptation. A basis for its non-redundant tangent space, which is at the same time a basis of symmetry eigenvectors, can only be constructed when the MPS reference is an eigenvector of those symmetries. If the MPS reference is a symmetry eigenvector, its tangent space (in general) also contains symmetry eigenvectors which belong to a different irreducible representation (irrep). We provide a simple counting argument.\\
Consider an MPS with length $L$ even, then the center virtual dimension has to be $d^{\frac{L}{2}}$ to represent a general FCI state.\cite{Schollwoeck} The number of states in its non-redundant tangent space is then
\begin{equation}
\text{dim} \mathbb{T} = \sum\limits_{k=1}^{\frac{L}{2}} \left( d d^k - d^{k-1} \right) d^{k-1} = d^L - 1,
\end{equation}
i.e. the rest of the Hilbert space. Note that these $d^L - 1$ non-redundant tangent space vectors can only be constructed if all Schmidt values are greater than zero.\cite{PhysRevLett.107.070601,2012arXiv1210.7710H} Suppose that this is the case. The MPS reference and its non-redundant tangent space then span the entire Hilbert space. If the MPS reference transforms according to a particular irrep of the symmetry group of the Hamiltonian, a basis of symmetry eigenvectors can be constructed for its non-redundant tangent space. If the Hilbert space is spanned by symmetry vectors belonging to at least two different irreps, the non-redundant tangent basis then contains symmetry eigenvectors belonging to a different irrep than the MPS reference.

\subsection{Implications for RPA}
If an MPS ansatz without symmetry adaptation is variationally optimized, it can occur that due to the choice of virtual dimensions a symmetry multiplet of a non-Abelian group (e.g. SU(2)) is incomplete at a certain boundary. From the projector interpretation of the non-redundant tangent space, it can be understood that this may lead to spurious zero energy RPA excitations: replacing an occuring renormalized basis state of the multiplet by one that is not in the renormalized basis, can lead to a state with the same energy and hence a spurious zero energy RPA excitation. For this reason, we have opted to use symmetry-adapted MPS references in this work.

\subsection{Tangent space of a symmetry-adapted ansatz}
We now discuss the construction of the tangent space of an SU(2) $\otimes$ U(1) adapted MPS ansatz. A spin- and particle number-adapted MPS decomposes into Clebsch-Gordan coefficients of the imposed symmetry groups and reduced tensors, due to the Wigner-Eckart theorem:\cite{wouters, *sharma:124121, 0295-5075-57-6-852, *1742-5468-2007-10-P10014, *PhysRevA.82.050301, *1367-2630-12-3-033029}
\begin{eqnarray}
& & A^{s s^z N}_{j_L j_L^z N_L \alpha_L ; j_R j_R^z N_R \alpha_R } \nonumber\\
& = & \braket{j_L j_L^z s s^z \mid j_R j_R^z} \delta_{N_L + N, N_R} T^{s N}_{j_L N_L \alpha_L ; j_R N_R \alpha_R }
\end{eqnarray}
The derivative operator $\frac{\partial}{\partial A^{\mu}}$ in Eq. \eqref{wavefunctionexpansion} is then replaced by $\frac{\partial}{\partial T^{\kappa}}$. All symmetry imposing Clebsch-Gordan coefficients are still in place, and the tangent space vectors hence have the same symmetry as the MPS reference. The non-redundant tangent space can be constructed in an analogous way as for the case without symmetry adaptation. The entire symmetry sector of the Hilbert space (minus the MPS reference) is retrieved in the non-redundant tangent space, if the virtual dimensions are taken as large as possible. The difference between the tangent spaces with and without symmetry-adaptation can be compared to the restricted and unrestricted HF manifolds.\cite{PhysRevA.22.2362,*PhysRevA.24.673} For the former only singlet excitations are possible, while for the latter triplet excitations are allowed too, even if the ground state is a singlet.\\
Note that if a symmetry-adapted MPS is optimized by the imaginary time-evolution of Sec. \ref{linktoTimeEvoMPS}, the distribution of the bond dimensions over the symmetry sectors is fixed. As such an optimization does not lead to an optimal distribution of the bond dimensions, we have used the two-site DMRG algorithm to optimize all the MPS reference wave-functions in this work.\\
Henceforth symmetry-adapted will be used as a shorthand for spin- and particle number-adapted.

\section{The 1D Hubbard chain}
In this section, we approximate low-lying eigenstates of the one-dimensional Hubbard chain with open boundary conditions (OBC) (see Eq. \eqref{HubbardModel}). The CISD-MPS ansatz of Eq. \eqref{CISDansatz}, which contains all excitations to the double tangent space, is used to improve on the ground state and to find low-lying excitations. The results are compared with TDA-MPS, which contains all excitations to the single tangent space. With RPA-MPS, we search for the Goldstone mode of a symmetry-broken ground state. In addition, we discuss RPA-MPS correlation energy calculations.
\subsection{CISD-MPS}
\begin{table*}[t]
\caption{\label{CISDtable}CISD-MPS improvement on the ground state and low-lying excitation energies for the one-dimensional Hubbard chain with length $L=8$ and OBC. D is the number of SU(2) multiplets retained at each boundary in the symmetry-adapted MPS reference calculation. The state for which the absolute energy is shown, was chosen as MPS reference. \\$\quad ^{(a)}$ Excitation with different multiplicity. The required FCI excitation is not in the TDA-MPS spectrum.}
\begin{tabular}{| c | c | c | c | c | c | c | c |}
\hline
U & Quantity & S & N & Exact (FCI) & TDA-MPS [D=3] & TDA-MPS [D=9] & CISD-MPS [D=3] \\
\hline
0.1 & E$_0$       & 0             & 8 & -9.319312 & -9.067465         & -9.301264 & -9.315185\\
    & E$_1$-E$_0$ & $\frac{1}{2}$ & 7 &  0.297631 &  0.222150         &  0.285466 &  0.311181\\
    & E$_2$-E$_0$ & $\frac{1}{2}$ & 9 &  0.397631 &  0.322150         &  0.385466 &  0.411181\\
    & E$_3$-E$_0$ & 0             & 6 &  0.611620 &  0.873285         &  0.619417 &  0.629720\\
\hline
1   & E$_0$-E$_1$ & $\frac{1}{2}$ & 7 & -0.022354 & -0.082237         & -0.029340 &  0.011799\\
    & E$_1$       & 0             & 6 & -7.790647 & -7.532068         & -7.780764 & -7.785715\\
    & E$_2$-E$_1$ & 0             & 8 &  0.095814 &  0.543100$^{(a)}$ &  0.105942 &  0.135785\\
    & E$_3$-E$_1$ & 1             & 6 &  0.517393 &  0.572255$^{(a)}$ &  0.530944 &  0.542513\\
\hline
10  & E$_0$       & 0             & 4 & -5.187427 & -5.083270         & -5.186955 & -5.187090\\
    & E$_1$-E$_0$ & $\frac{1}{2}$ & 5 &  0.008950 & -0.010314         &  0.009270 &  0.010988\\
    & E$_2$-E$_0$ & 1             & 4 &  0.113988 &  0.127636         &  0.114721 &  0.114984\\
    & E$_3$-E$_0$ & $\frac{1}{2}$ & 5 &  0.189005 &  0.205577$^{(a)}$ &  0.196828 &  0.192880\\
\hline
100 & E$_0$       & 0             & 4 & -4.805753 & -4.736845         & -4.805615 & -4.805360\\
    & E$_1$-E$_0$ & 1             & 4 &  0.013020 &  0.013672         &  0.013016 &  0.013783\\
    & E$_2$-E$_0$ & 1             & 4 &  0.027045 &  0.022700         &  0.027013 &  0.028034\\
    & E$_3$-E$_0$ & 0             & 4 &  0.034327 &  0.316707$^{(a)}$ &  0.034327 &  0.035940\\
\hline
\end{tabular}
\end{table*}
The TDA and CISD calculations were done by optimizing a symmetry-adapted MPS reference, with $D$ retained multiplets at each boundary. This reference was then used in TDA and CISD calculations without symmetry constraints. As the symmetry-adapted reference is not necessarily a variational minimum for an MPS ansatz without symmetry constraints, negative excitation energies can occur.\\
The CISD ansatz in Eq. \eqref{CISDansatz} leads to a generalized eigenvalue problem
\begin{equation}
\braket{\overline{\Phi^0_{\kappa \lambda}} \mid \hat{H} \mid \Phi^0_{\mu \nu} } C^{\mu \nu} = E \braket{\overline{\Phi^0_{\kappa \lambda}} \mid \Phi^0_{\mu \nu} } C^{\mu \nu}
\end{equation}
which was solved by multi-targeting the lowest states with the Davidson algorithm.\cite{Davidson197587} By decomposing the $C$-tensor, the CISD ansatz can be written as a sum over MPS wave-functions:
\begin{equation}
\ket{\text{CISD}} = \sum\limits_{i < j} C^{\mu_i \nu_j} \ket{\Phi^0_{\mu_i \nu_j}} = \sum\limits_{i < j} \sum\limits_{p(i,j)} C^{\mu_i}_{L,p} C^{\nu_j}_{R,p} \ket{\Phi^0_{\mu_i \nu_j}}. \label{sumoverMPSs}
\end{equation}
This allows to use standard MPS machinery\cite{Schollwoeck} in the matrix-vector multiplication. Because the sum of several MPS wave-functions yields an MPS with a larger bond dimension,\cite{Schollwoeck} this immediately leads to the understanding that the CISD ansatz can introduce extra entanglement.\\
We chose $L=8$ and four $U$-values: $0.1$, $1$, $10$ and $100$. With increasing $U$, the ground state changes from a collection of quasi-free particles to a highly correlated state. For the latter, HF gives a qualitatively wrong description. For $U=1$, the ground state contains $7$ particles and has spin $\frac{1}{2}$. If a symmetry-broken reference is chosen, the multiplet degeneracy of the excitations is lost. Therefore, we opted for the first singlet state as MPS reference for $U=1$. Although the TDA and CISD calculations were not symmetry-adapted, the multiplet degeneracy was \textit{exactly} retrieved, because we started from a symmetry-adapted MPS reference. The first four multiplets for each U-value are shown in Table \ref{CISDtable}.\\
The TDA-MPS [D=9] energies and the CISD-MPS [D=3] energies are of roughly the same quality, and improve on the TDA-MPS [D=3] energies significantly. The CISD-MPS ansatz contains $\mathcal{O}(\frac{1}{2} (L d D^2)^2)$ variational parameters to include extra correlation between all pairs of sites on top of the MPS reference, and can be used both to improve on the ground state and to approximate excited states. The TDA-MPS ansatz contains $\mathcal{O}(L d D^2)$ variational parameters, and due to the MPS analogue of Brillouin's theorem, it can only be used to approximate excited states. The relative accuracy of the CISD-MPS [D=3] and MPS [D=9] reference state energies changes with $U$. With increasing single-particle nature, the CISD-MPS ansatz performs better for the targeted reference.\\
For small $D$, not all excited states are retrieved with TDA-MPS. An example is the third excited state for D=3 and U=100. The targeted state consists of two singlet-triplet excitations, which interact to form a bound singlet state. This is well captured by CISD-MPS [D=3] and TDA-MPS [D=9], while $E_3$ for TDA-MPS [D=3] in Table \ref{CISDtable} corresponds in fact to a doublet. The MPS [D=9] reference has a large enough bond dimension to capture the two excitations in its single tangent space, while the CISD-MPS [D=3] ansatz captures the double excitation in the MPS's double tangent space.

\subsection{RPA-MPS and Goldstone modes}
The $L=8$ and $U=1$ case is an ideal candidate to retrieve a Goldstone mode, because a specific spin-$\frac{1}{2}$ ground state vector is necessarily a symmetry-broken state. With an MPS reference optimized without any symmetry constraints and $D=16$ (now exceptionally the number of states instead of the number of multiplets), we find one zero energy solution to the RPA equations, and this solution also satisfies Eq. \eqref{GoldstoneEq}. This is the Goldstone mode from the symmetry-broken spin-$\frac{1}{2}$ ground state. Zero energy solutions can also arise for singlet ground states, if the MPS accidently breaks non-Abelian symmetries, as discussed in Sec. \ref{SyAdaptSec}. This can be avoided by retaining complete multiplets at each boundary, whereas the RPA Goldstone mode for a symmetry-broken ground state will always occur, even for bond dimensions that reproduce the full Hilbert space.

\subsection{The RPA-MPS correlation energy}
\begin{figure}[t]
\includegraphics[width=0.45\textwidth]{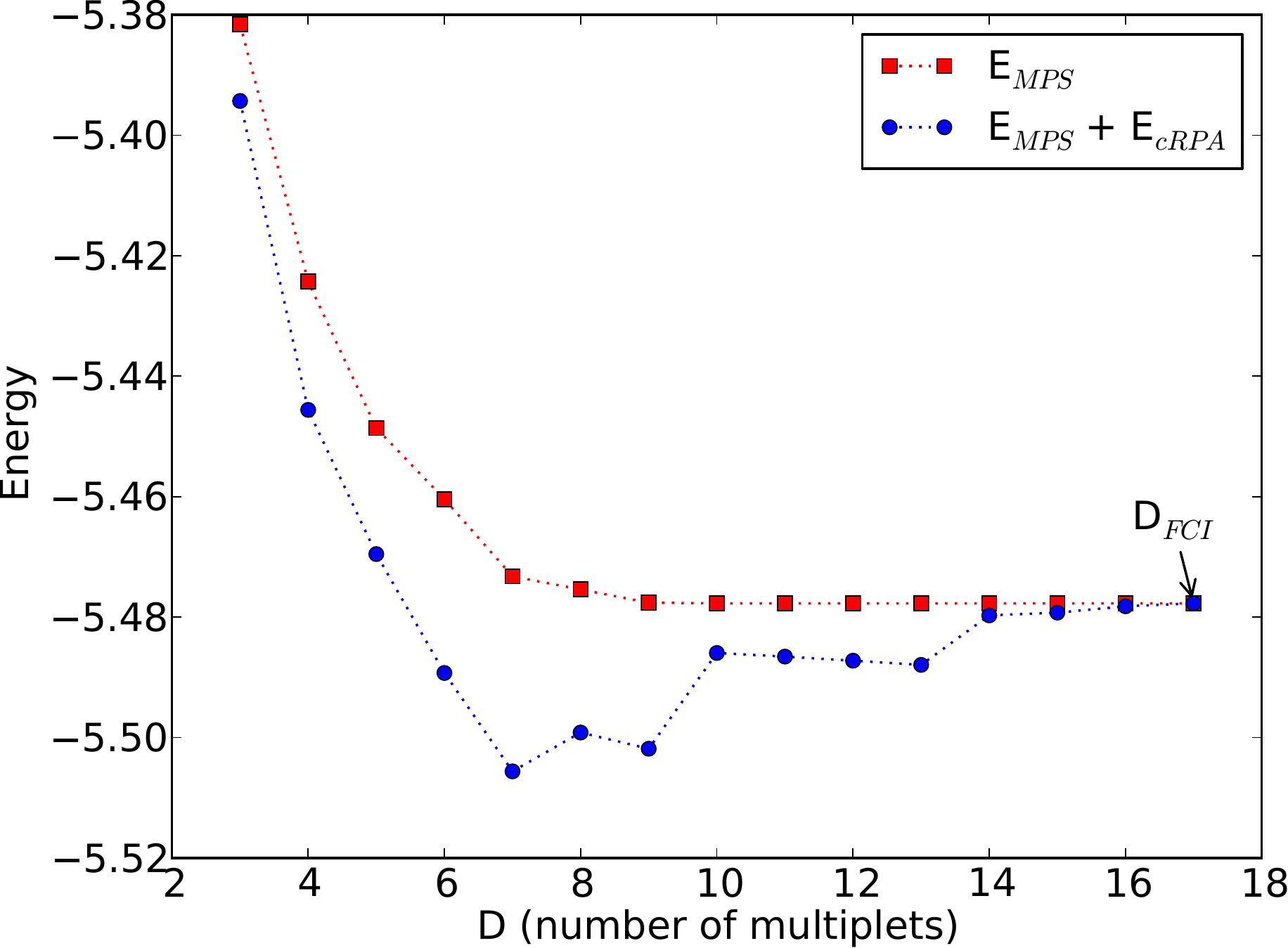}
\caption{\label{RPAcorrFig} The RPA-MPS correlation energy for the Hubbard chain with OBC, length $L=6$, filled with $N=4$ electrons, in the singlet state, and $U=1$. A symmetry-adapted ansatz was used, with D retained multiplets at each boundary.
}
\end{figure}
We calculated RPA-MPS correlation energies for symmetry-adapted ansatzes. Remember that only excitations with the same symmetry as the MPS reference are then retrieved. For the Hubbard chain with OBC and length $L=6$, filled with $N=4$ electrons, in the singlet state, and $U=1$, $E_{\text{cRPA}}$ is shown in Fig. \ref{RPAcorrFig}. From Eq. \eqref{RPAcorrEnergyEq2}, $2 E_{\text{cRPA}}$ can be interpreted as the mean difference between RPA and TDA excitation energies, multiplied by the number of excitations ($\text{dim} \mathbb{T}$). With increasing $D$, $| E_{\text{cRPA}} |$ first increases because the number of excitations increases. For even larger $D$, the mean difference between the RPA and TDA energies vanishes faster than the number of excitations increases. When the FCI virtual dimensions are reached, the RPA and TDA excitation energies are equal, as the $B$-matrix vanishes, and $E_{\text{cRPA}}$ is exactly zero.\\
When calculating $E_{\text{cRPA}}$, care has to be taken that the MPS reference is the variational minimum and that no symmetries are broken, such as the multiplet structure at a boundary or e.g. the SO(4) symmetry when considering a half-filled Hubbard chain.\cite{doi:10.1142/S0217984990000933} If these conditions are not fulfilled, the RPA-MPS correlation energy breaks down ($| E_{\text{cRPA}} | \gg | E_{\text{MPS}} |$).\\

\section{Polyenes}
\begin{figure}[t]
\includegraphics[width=0.45\textwidth]{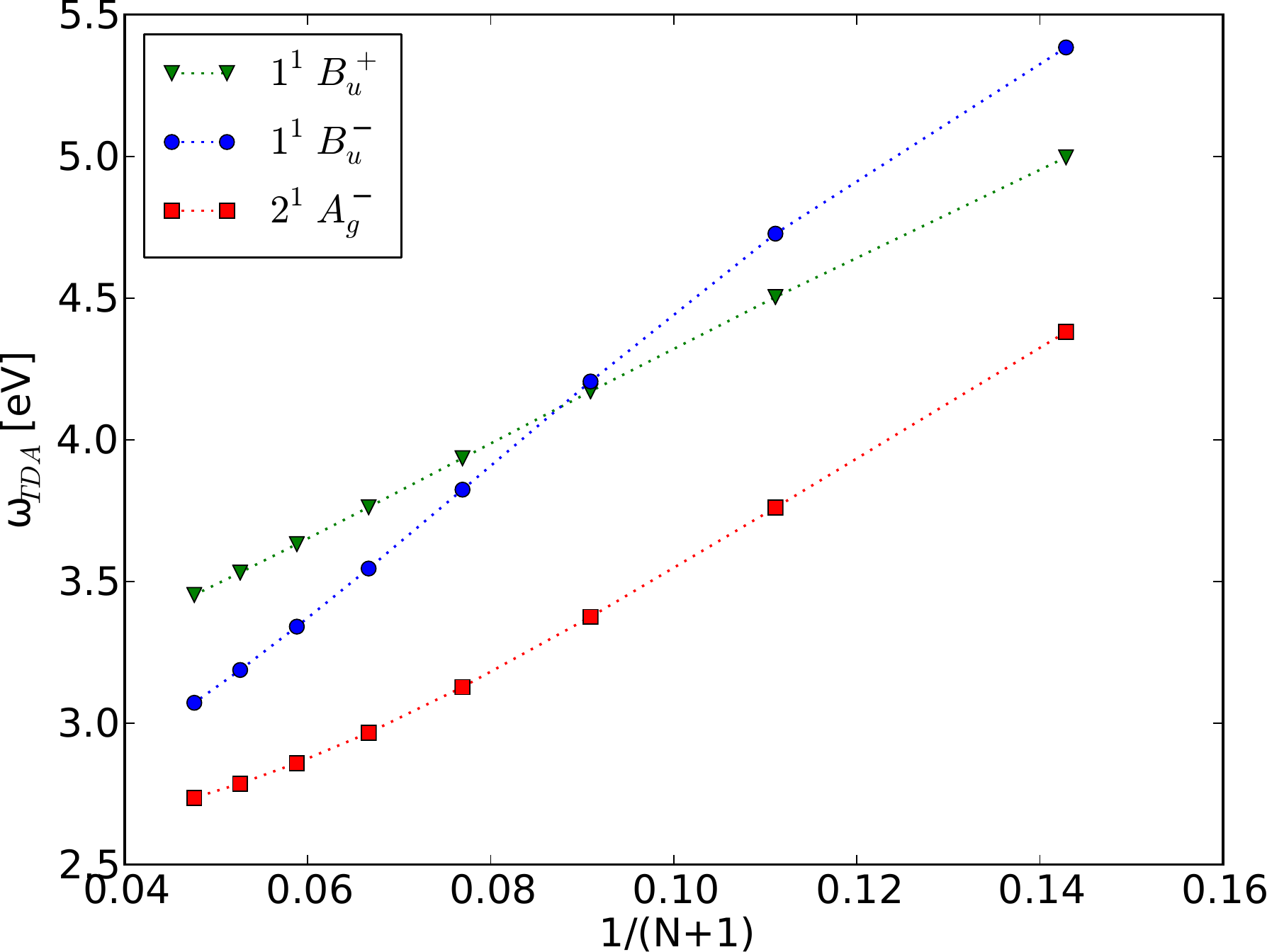}
\caption{\label{PPPTDA} The first three TDA-MPS excitation energies for a polyene chain with $N$ carbon atoms, for which the $\pi$-system was approximated by the long-range PPP Hamiltonian.}
\end{figure}

\begin{figure}[b]
\includegraphics[width=0.45\textwidth]{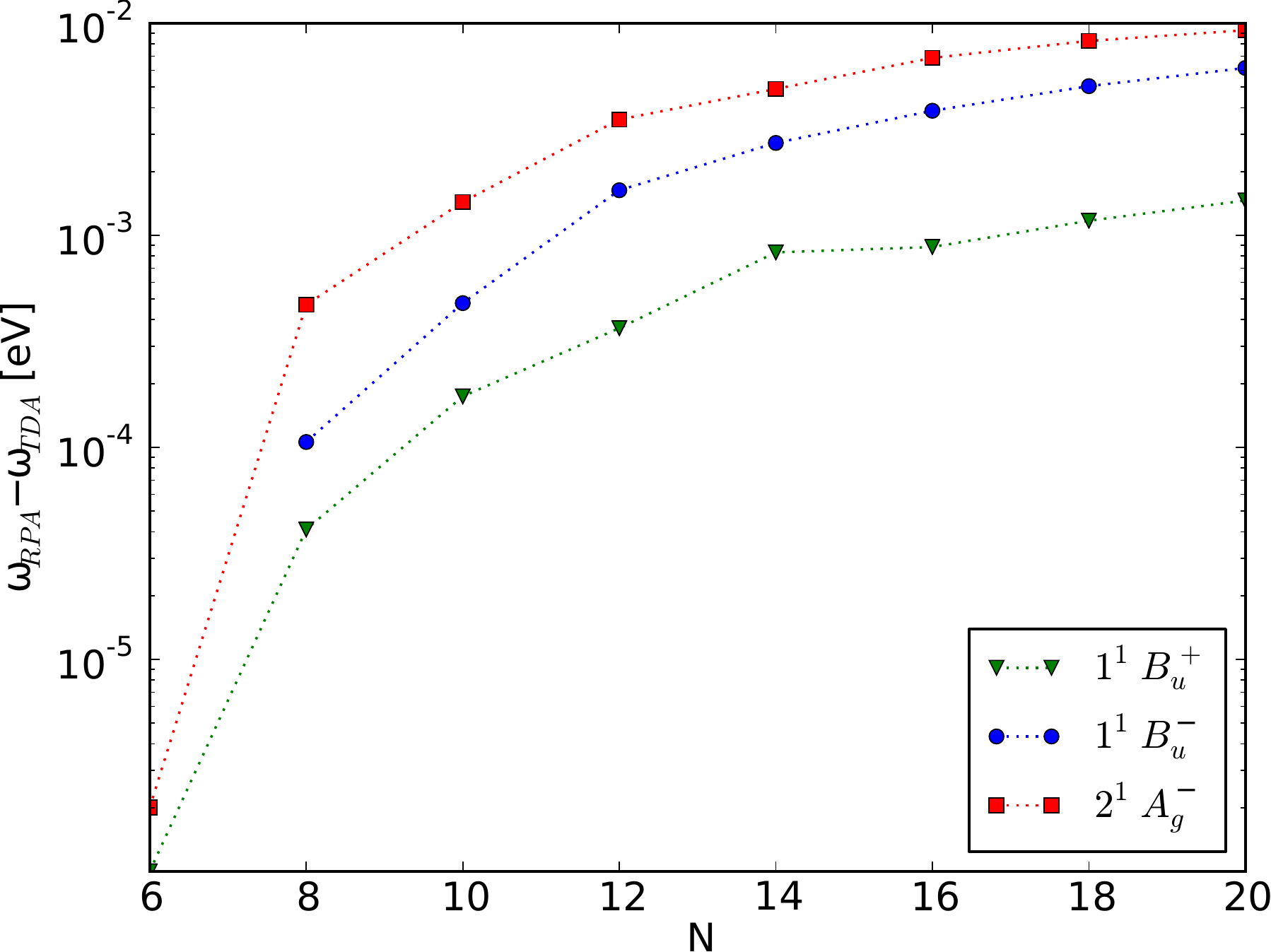}
\caption{\label{PPPTDARPAfig} The difference between the RPA-MPS and TDA-MPS energies for the first three excitations of a polyene chain with $N$ carbon atoms, for which the $\pi$-system was approximated by the long-range PPP Hamiltonian.}
\end{figure}

Polyenes are linear conjugated chains of hydrocarbons:
\begin{equation}
\text{CH}_2 = \text{CH} - \text{CH} = \text{CH} - \text{CH} = \text{CH}_2.
\end{equation}
Excitations in the $\pi$-system lie in the visible region of the spectrum, and polyenes are therefore important building blocks for light absorption and dyes. They have a long history of use as benchmark systems to test new quantum chemistry excited state methods. The $\pi$-system can be approximated by the long-range Pariser-Parr-Pople (PPP) Hamiltonian, where the two-body terms of the Hamiltonian in Eq. \eqref{HamiltonianSecondQuantization} are approximated by a local Coulomb repulsion:
\begin{equation}
\hat{V} = \frac{1}{2} \sum\limits_{k l \sigma \tau} R^{kl} \hat{n}_{k \sigma} \hat{n}_{l \tau}.
\end{equation}
The Latin letters denote orbitals and the Greek letters spin projections. For our calculations, we used the Ohno parametrization for the electron-electron repulsion $R^{kl}$.\cite{Ohno} All Hamiltonian parameters, except $R_{^{\text{single}}_{\text{double}}} = 1.40 \pm 0.05 \AA$, are identical to the ones in Ref. \onlinecite{ohmine:2298}.\\
Many DMRG calculations studying the excited states and response properties of conjugated molecules have been performed, using a parametrized Hamiltonian.\cite{PhysRevB.54.7598, *PhysRevB.55.15368, *PhysRevB.56.9298, *anusooya:10230, *PhysRevB.66.035116, *mukhopadhyay:074111} At the ab initio level, high-lying excited states have been targeted with the harmonic Davidson adaptation of the DMRG method.\cite{dorando:084109} Frequency-dependent dipole polarizabilities were computed at the ab initio level by \textcite{dorando:184111} using the TDA-MPS approximation.\\
Using the PPP Hamiltonian, we approximated the first three particle-conserving singlet excitations with the symmetry-adapted RPA-MPS and TDA-MPS methods. We kept $D=20$ retained multiplets at each boundary. The TDA-MPS excitation energies are shown in Fig. \ref{PPPTDA} as a function of the number of carbon atoms $N$ in the polyene. The symmetry labeling was based on Fig. 10 in Ref. \onlinecite{PhysRevB.36.4337}. The difference between the RPA-MPS and TDA-MPS energies is shown in Fig. \ref{PPPTDARPAfig}, indicating that the ground state MPS reference is already quite accurate for $D=20$, as the B-matrix contributions are small. The RPA-MPS and TDA-MPS excitation energies match better for the higher excitations of Fig. \ref{PPPTDA}.


\section{Summary}
In this work, we attempted to set up a post-DMRG framework by finding the excitation structure of the MPS reference. As a guide, we carefully followed the structure of HF theory and the subsequent post-HF methods, exploiting the fact that both HF and DMRG can be seen as product-like wave-functions.\cite{B805292C}\\
A variational wave-function ansatz can be used in the TIVP to yield self-consistent equations. With the TDVP, optimal time-evolution is found which stays within the ansatz manifold. Linearization of the TDVP around a variational minimum gives the RPA equations. The optimal time-evolution requires a non-redundant parametrization of the ansatz's tangent space to exclude meaningless variations of the wave-function. Occupied-occupied variations in HF theory, as well as variations in the direction of the renormalized DMRG basis states, only lead to norm or phase changes. They do not change the physical state represented by the ansatz.\\
Exponentiation of the norm- and phase-conserving variations in HF theory, led to the Thouless theorem: a non-redundant parametrization of the entire HF (Grassmann) manifold, generated by the OV excitations of any particular SD. In this work, we have proposed the DMRG counterpart: a non-redundant parametrization of the entire MPS manifold, generated by the norm- and phase-conserving changes of any particular MPS wave-function. Just like the norm- and phase-conserving changes of HF theory are generated by replacing an occupied orbital by a virtual orbital, the norm- and phase-conserving changes of an MPS wave-function are generated by replacing the occuring renormalized basis states by discarded renormalized basis states. We have proven the MPS counterpart of Thouless's theorem for a general MPS with OBC, for which no Schmidt values vanish.\\
By identifying the excitation structure of the SD/MPS ansatz by means of the Thouless theorem, the RPA equations can be rederived be means of the EOM. This allows for a bosonic expansion of the Hamiltonian, and the definition of the RPA correlation energy and wave-function.\\
The different orders of tangent space of the Thouless parametrization generate the CI basis. Eigenstates of the Hamiltonian can be approximated in this basis. CIS, or CI with only single excitations, yields again the SD/MPS reference due to Brillouin's theorem, as well as a set of excited states. These excited states are found by diagonalizing the Hamiltonian in the non-redundant tangent basis, or the A-matrix of RPA. This method is known as TDA.\\
When the MPS reference is a good approximation of the true ground state, $\| \left( \hat{H} - E_{\text{MPS}} \right) \ket{\Phi^0} \|_2$ becomes small, and the B-matrix contributions of RPA vanish. TDA and RPA then lead to the same excitation energies.\\ The RPA wave-function suggests a size-consistent CC ansatz on top of the reference wave-function.\\
The ideas presented in this paper are illustrated with proof-of-principle calculations of CISD-MPS improvements on the ground state, TDA-MPS, RPA-MPS and CISD-MPS excitation energies, an RPA-MPS Goldstone mode, and the RPA-MPS correlation energy. In contrast to HF, the MPS reference gives also in the highly correlated regime of the Hubbard model a qualitatively good description, and variational post-DMRG methods such as TDA-MPS and CISD-MPS give numerically relevant results. For an MPS with small bond dimensions, two correlated single excitations are not always retrieved in the tangent space, and the CISD-MPS ansatz is a better choice then.\\

Near completion of this work, we learned about Ref. \onlinecite{JuthoSimul} which presents RPA-MPS calculations and new multi-site excitation ansatzes for uniform MPS.

\begin{acknowledgments}
This research was supported by the Research Foundation Flanders (S.W.) and the National Science Foundation grant no. SI2-SSE:1265277 (G.C.). The authors would like to thank Jutho Haegeman, Frank Verstraete and Stijn De Baerdemacker for the many stimulating conversations.
\end{acknowledgments}

\appendix*
\section{Explicit Grassmann manifold parametrization \label{appendixSec}}
The proof given here is inspired by the proof for the unitary counterpart of Thouless's theorem for HF, given in \textcite{PhysRevA.22.2362} Give a unitary $m \times n$ matrix $Q$ with $m>n$, i.e. $Q^{\dagger}Q = I_n$, and a second unitary matrix $U$, of the same form as $Q$. Form a unitary $m\times (m-n)$ matrix $\tilde{Q}$ so that $[Q \tilde{Q}] [Q \tilde{Q}]^{\dagger} = I = [Q \tilde{Q}]^{\dagger} [Q \tilde{Q}]$. For the parametrization
\begin{equation}
\tilde{A}(\mathbf{y},\overline{\mathbf{y}}) = \exp{\left( \tilde{Q} y Q^{\dagger} - Q y^{\dagger} \tilde{Q}^{\dagger} \right)} Q \label{appendixEq}
\end{equation}
with $y$ an $(m-n) \times n$ matrix containing the complex variables and $\mathbf{y}$ the corresponding flattened column, there exists at least one $\mathbf{y_u}$ so that the columns of $\tilde{A}(\mathbf{y_u},\overline{\mathbf{y_u}})$ and the columns of $U$ (denoted by $\mathbf{u_k}$) span the same space. We will provide a proof by construction:
\begin{enumerate}

\item The matrix $M = U^{\dagger} Q Q^{\dagger} U$ is a hermitian positive semidefinite matrix. There exists a unitary transformation to rotate the basis $\mathbf{u_i}$ to $\mathbf{v_i}$, so that $\mathbf{v_k}^{\dagger} Q Q^{\dagger} \mathbf{v_j} = \delta_{kj} n_k^2$.

\item Write $\mathbf{v_i}$ in terms of $\mathbf{q_k}$ and $\tilde{\mathbf{q}}_\mathbf{k}$, the columns of $Q$ and $\tilde{Q}$: $\mathbf{v_i} = \sum_k \alpha_{ik} \mathbf{q_k} + \sum_l \beta_{il} \tilde{\mathbf{q}}_\mathbf{l}$. From the previous step we know that $\delta_{ij} = \mathbf{v_i}^{\dagger} \mathbf{v_j} = n_i^2 \delta_{ij} + \sum_l \overline{\beta_{il}} \beta_{jl}$.

\item If $n_i \neq 0$, define $\mathbf{r_i}$ by $n_i \mathbf{r_i} = \sum_k \alpha_{ik} \mathbf{q_k}$. If $n_i \neq 1$, define $\tilde{\mathbf{r}}_{\mathbf{i}}$ by $(1 - n_i^2)^{\frac{1}{2}} \tilde{\mathbf{r}}_{\mathbf{i}} = \sum_l \beta_{il} \tilde{\mathbf{q}}_{\mathbf{l}}$. Note that if e.g. $2n>m$, a number of $n_i$ will certainly be 1, and the corresponding vectors $\tilde{\mathbf{r}}_{\mathbf{i}}$ cannot be constructed.

\item From the previous steps, it follows that the vectors $\{ \mathbf{r_i}$, $\tilde{\mathbf{r}}_{\mathbf{i}} \}$ are orthonormal. Complete both sets with additional vectors, so that they span the same space as the columns of resp. $Q$ and $\tilde{Q}$. If the matrix $R$ contains $\mathbf{r_i}$ in its columns and $\tilde{R}$ contains $\tilde{\mathbf{r}}_{\mathbf{i}}$ in its columns, a unitary transformation $P$ links $R$ and $Q$ by $R = QP$ and a unitary transformation $\tilde{P}$ links $\tilde{R}$ and $\tilde{Q}$ by $\tilde{R} = \tilde{Q} \tilde{P}$.

\item If $n_i \neq 1$, consider $\mathbf{w_i} = \exp{\left( \gamma_i ( \tilde{\mathbf{r}}_{\mathbf{i}} \mathbf{r_i}^{\dagger} - \mathbf{r_i} \tilde{\mathbf{r}}_{\mathbf{i}}^{\dagger} ) \right)} \mathbf{r_i} = \cos{(\gamma_i)} \mathbf{r_i} + \sin{(\gamma_i)} \tilde{\mathbf{r}}_{\mathbf{i}}$. Assign $0 \leq \gamma_i \leq \frac{\pi}{2}$ so that $\cos{(\gamma_i)} = n_i$. It then follows that $\mathbf{w_i} = \mathbf{v_i}$. Note that if $n_i=1$, $\gamma_i$ would have been 0, and that the exponential in front of $\mathbf{r_i}$ then becomes the identity. So it poses no problem that the corresponding vectors $\tilde{\mathbf{r}}_{\mathbf{i}}$ cannot be constructed.

\item If $\gamma$ is regarded as a diagonal matrix containing the $\gamma_i$ values, the singular value decomposition of $y_u$ is given by $y_u = \tilde{P} \gamma P^{\dagger}$. This can be confirmed by writing the exponential expression for $\mathbf{w_i}$ in terms of $Q$ and $\tilde{Q}$.

\end{enumerate}
This concludes the construction of the complex $(m-n) \times n$ matrix $y_u$. Eq. \eqref{appendixEq} hence represents a Grassmann manifold.

\bibliographystyle{apsrev4-1}
\bibliography{biblio}

\end{document}